\documentclass[useAMS,usenatbib]{mnras}%
\usepackage{courier}
\usepackage[T1]{fontenc} 
\usepackage[utf8]{inputenc}
\usepackage{ae,aecompl}
\usepackage{times}
\usepackage{amsmath}
\usepackage{amssymb}
\usepackage[english]{babel}
\usepackage{graphicx}
\usepackage[percent]{overpic}
\usepackage{tikz}
\usetikzlibrary{decorations.pathreplacing,external}
\usepackage{pdfpages}
\usepackage{flafter}
\usepackage{multirow}
\usetikzlibrary{arrows}
\usepackage{booktabs}
\usepackage[]{natbib}
\usepackage{color}
\hypersetup{draft}
\usepackage{gensymb}

\newcommand{\ud}{d}

\newcommand{\eq}[1]{Eq.~(\ref{#1})}
\newcommand{\fig}[1]{Fig.~\ref{#1}}
\newcommand{\msun}{{\ensuremath{M_\odot}}}

\def\ie{{\it i.e.,}\,}
\def\eg{{\it e.g.,}\,}

\title[Gravitational Waves]{Gravitational waves from three-dimensional core-collapse supernova models: 
The impact of moderate progenitor rotation}
\author[H.~Andresen et al.]{
H.~Andresen$^{1,2,3}$\thanks{E-mail: haakon.andresen@aei.mpg.de},
E.~M\"uller$^{1}$, H.-Th.~Janka$^{1}$, A.~Summa$^{1}$, K.~Gill$^{4,5}$, and M.~Zanolin$^{6}$\\
$^{1}$Max-Planck-Institut f\"ur Astrophysik,Karl-Schwarzschild-Str. 1, D-85748 Garching, Germany
\\
$^2$Physik Department, Technische Universit\"at M\"unchen, James-Franck-Str. 1, 85748 Garching, Germany
\\
$^3$
Max Planck Institute for Gravitational Physics (Albert Einstein  Institute), Am  M\"uhlenberg  1,  Potsdam-Golm, 14476, Germany
\\
$^4$Harvard-Smithsonian Center for Astrophysics, 60 Garden Street, Cambridge, Massachusetts 02138, USA 
\\
$^5$Department of Physics, Columbia University, New York, New York 10027, USA
\\
$^6$Embry Riddle University, 3700 Willow Creek Road, Prescott, Arizona 86301, USA
}
\def\LaTeX{L\kern-.36em\raise.3ex\hbox{a}\kern-.15em
           T\kern-.1667em\lower.7ex\hbox{E}\kern-.125emX}

\begin{document}
\maketitle

\begin{abstract}
We present predictions for the gravitational-wave (GW) emission
of three-dimensional supernova (SN) simulations performed
for a 15 solar-mass progenitor with the \textsc{Prometheus-Vertex}
code using energy-dependent, three-flavor neutrino transport. 
The progenitor adopted from stellar evolution calculations including
magnetic fields had a fairly low specific angular momentum 
($j_\mathrm{Fe} \lesssim 10^{15}$\,cm$^2$s$^{-1}$) in the iron core 
(central angular velocity $\Omega_\mathrm{Fe,c}\sim$0.2\,rad\,s$^{-1}$), 
which we compared to simulations without rotation and with 
artificially enhanced rotation 
($j_\mathrm{Fe} \lesssim 2\times  10^{16}$\,cm$^2$s$^{-1}$;
$\Omega_\mathrm{Fe,c}\sim$0.5\,rad\,s$^{-1}$). Our results 
confirm that the time-domain GW signals of SNe are stochastic,
but possess deterministic components with characteristic patterns
at low frequencies ($\lesssim$200\,Hz), caused by mass motions 
due to the standing accretion shock instability (SASI), and at
high frequencies, associated with gravity-mode oscillations in
the surface layer of the proto-neutron star (PNS). Nonradial mass
motions in the postshock layer as well as PNS convection are important
triggers of GW emission, whose amplitude scales with the power
of the hydrodynamic flows. There is no monotonic increase of the GW
amplitude with rotation, but a clear correlation with the strength 
of SASI activity. Our slowly rotating model is a fainter GW emitter
than the nonrotating model because of weaker SASI
activity and damped convection in the postshock layer and PNS.
In contrast, the faster rotating model exhibits a powerful SASI spiral
mode during its transition to explosion, producing the highest GW
amplitudes with a distinctive drift of the low-frequency emission 
peak from $\sim$80--100\,Hz to $\sim$40--50\,Hz. This migration
signifies shock expansion, whereas non-exploding models are 
discriminated by the opposite trend.
\end{abstract}

\section{Introduction}
Gavitational waves (GW) are generated in core-collapse supernovae by
time-dependent rotational flattening particularly during collapse and
bounce, prompt post-shock convection, non-radial flow inside the
proto-neutron star and in the neutrino-heated hot bubble, the activity
of the standing accretion shock instability (SASI), asymmetric
emission of neutrinos, and by asymmetries associated with the effects
of magnetic fields \citep[for recent reviews see, \eg][]{FryerNew11,
  Kotake13, MuellerB17}.  Measurable impact of rotation on the GW
signature is only expected for particular progenitors that possess a
sufficient amout of angular momentum, while all other processes are
genuinely operative in any core-collapse supernova.

Although, the rotation rate of most core-collapse progenitors might be
slow rather than rapid \citep{heger_05, beck_12, mosser_12, popov_12,
  noutsos_13, cantiello_14, deheuvels_14}, the study of GW generated
by the core-collapse of rotating stars has a long and rich history
\citep[for a review see, \eg][]{FryerNew11, Kotake13}.  Most earlier
studies of the GW signature of core-collapse supernovae were concerned
only with the collapse, bounce, and early ($\la 20$\,ms) post-bounce
evolution of a rotating iron core assuming axisymmetry \citep[see,
\eg][]{mueller_82, moenchmeyer_91, zwerger_97, dimmelmeier_02_b,
  kotake_03, shibata_04, dimmelmeier_07_a, dimmelmeier_08} or none
\citep[see, \eg][]{rampp_98, shibata_05, Ott+07, Scheidegger+10}.
Studies of the influence of rotation (and of a magnetic field) on the
GW signal from the phase of neutrino-driven convection and SASI, and
the onset of explosion or black hole formation have become available
more recently \citep{ott_11, kotake_11, ott_12, CerdaDuran+13,
  kuroda_14, Yokozawa+15, hayama_16, TakiwakiKotake18}. They showed
that a dominant source of post-bounce GW emission in rotating cores is
due to non-axisymmetric instabilities, which also have an important
influence on the explosion \citep{kotake_11, kuroda_14, takiwaki_16,
  TakiwakiKotake18}. All of these 3D GW signal predictions rely on a
simplified treatment of neutrino transport. Up to now, only
\citet{Summa_18} have provided self-consistent 3D models for a
rotating progenitor, but the GW signature of these models will only
be discussed in this paper.

For a more rapidly rotating case than considered by \cite{Summa_18}, the
post-bounce GW signal shows clear imprints of non-axisymmetric
instabilities arising from the low-T/W instability
\citep{takiwaki_16, TakiwakiKotake18}.
This 3D instability leads to a pole-to-equator contrast in the GW
amplitudes.  The emission, peaking at about 240\,Hz, is stronger along
the spin axis of the rotating core, and the amplitude contrast is
larger for the $h_x$ polarization mode than for the $h_+$ mode
\citep{TakiwakiKotake18}.  This result confirms earlier findings which
were obtained in simulations using a more approximate neutrino
transport \citep{Ott09, Scheidegger+10}. The GW frequency is twice the
modulation frequency of the neutrino signal, and hence joint GW and
neutrino observations could provide evidence for or against rapid core
rotation \citep{Ott+12, Yokozawa+15, kuroda_17, TakiwakiKotake18}.
These GW signal predictions were obtained, however, from 3D models
with an initial pre-collapse angular frequency of $\Omega_0 = 2$ rad/s in the
iron core and with simplified treatment of neutrino transport.

Here, we present the GW signals from three 3D core-collapse
simulations performed by \citet{Summa_18} to study the possible
support of neutrino-driven supernova explosions by more modest rotation.  These
self-consistent 3D simulations applied the most complete set of
neutrino interactions currently available, and are based on the
spherically symmetric $15 \msun$ progenitor model m15u6 from stellar
evolution calculations by \citet{heger_05}
\footnote{http://www.2sn.org/stellarevolution/magnet/}
, which was evolved in 1D including the effects of rotation and angular
momentum transport by magnetic fields. \citet{Summa_18} also performed
two additional core-collapse simulations for the same progenitor.
In one model the rate of rotation was increasing to $\Omega_0 = 0.5$ rad/s
in the central iron core. In the second additional simulation the
initial rotation rate was set to zero throughout the stellar progenitor.
model. We have also evaluated the GW signal of these two models.
The 3D simulations of \citet{Summa_18} start only approximately 10\,ms
after core bounce (the earlier evolution was simulated in
axisymmetry), because deviations from axisymmetry should not occur
until this time for the moderately fast rotating progenitors they
adopted for their work. Hence, the expected GW signal associated with
core bounce of the two rotating models was obtained from axisymmetric
configurations. Analyzing the GW signature of the three models allows
us study how the signal changes, for the same stellar progenitor, as a
function of progenitor rotation.  The fastest rotating model and the
model without rotation are dominated by strong SASI activity, while
the slowly rotating model develops only weak and intermittent SASI
oscillations, \ie we can also ascertain the influence of rotation in
both the SASI dominated regime and the convective regime on the GW
signal.

The paper is structured as follows: First, we recap the numerical
methods and input physics used in the code of \citet{Summa_18} to
perform the numerical simulations of the 3D supernova models. In
section~\ref{sec:mod}, we describe the relevant properties of the
models of \citet{Summa_18} that we have analyzed for their GW
signature, and we briefly discuss their dynamics. The formalism we
used to extract the GW signature from the models of \citet{Summa_18}
is described in section~\ref{sec:gw}.  In sections~\ref{sec:res},
\ref{sec:p2ext}, \ref{sec:sasi}, and \ref{sec:cb} we present the GW signals,
discussing the underlying hydrodynamic effects responsible for GW
excitation and how these effects are affected by rotation.
We briefly discuss the GW signal associated with core bounce (simulated only in 2D; see
above) for completeness.

Before we
give our conclusions in section~\ref{sec:con}, we assess the detection
prospects for the GW signals of the three models analyzed by us in
section~\ref{sec:det}.

%%%%%%%%%%%%%%%%%%%%%%%%%%%%%%%%%%%%%%%%%%%%%%%%%%%%%%%%%%%%%%%%%%%%%%%%%%%%
\section{Numerical Methods and Input Physics}
The \textsc{Prometheus-Vertex} code \citep{rampp_02,buras_06a} was
used to perform the core-collapse simulations of \citet{Summa_18}.
\textsc{Prometheus-Vertex} consists of two main modules. The
hydrodynamics calculations are handled by \textsc{Prometheus}
\citep{mueller_91,fryxell_91}, which is a Newtonian hydro-code that
implements the piecewise parabolic method of \cite{colella_84} in
spherical coordinates $(r,\theta,\phi)$.  The monopole approximation
is used to treat self-gravity and general relativistic effects are
taken into account by means of a pseudo-relativistic potential (case~A
of \citealt{marek_06}).  The neutrino transport is taken care of by
the module \textsc{Vertex} \citep{rampp_02}, which solves the
energy-dependent two-moment equations for three neutrino species,
electron neutrinos ($\nu_e$), anti-electron neutrinos ($\bar{\nu}_e$),
and a third species ($\nu_X$) representing all the heavy flavor
neutrinos, using a variable Eddington-factor closure computed from solutions of
the Boltzmann equation. Multi-dimensional transport is approximated by the
``ray-by-ray-plus'' method  of \citet{buras_06a}.
The high-density equation of state (EoS) used is
the nuclear equation of state (EoS) of \citet{lattimer_91}, with
$K=220 \,\mathrm{MeV}$. Below $10^{11}$g/cm$^3$ a low-density EoS
for nuclei, nucleons, charged leptons and photons is applied.

%%%%%%%%%%%%%%%%%%%%%%%%%%%%%%%%%%%%%%%%%%%%%%%%%%%%%%%%%%%%%%%%%%%%%%%%%%%%
\section{Supernova Models} 
\label{sec:mod}
We extracted the GW signal of three models of \citet{Summa_18} based
on the progenitor of \cite{heger_05}, which is a solar-metallicity
star with a zero-age main sequence (ZAMS) mass of $15$ solar masses
($\msun$).  The stellar evolution calculation of \citet{heger_05}
accounted for the effects of magnetic fields and rotation, they
evolved the model from the ZAMS to the onset of core-collapse. The
inclusion of magnetic fields leads to a dramatic overall reduction of the
final rotation rate of the iron core, compared to calculations without
magnetic fields.
\begin{figure}           
\centering                            
\includegraphics[width=0.45\textwidth]{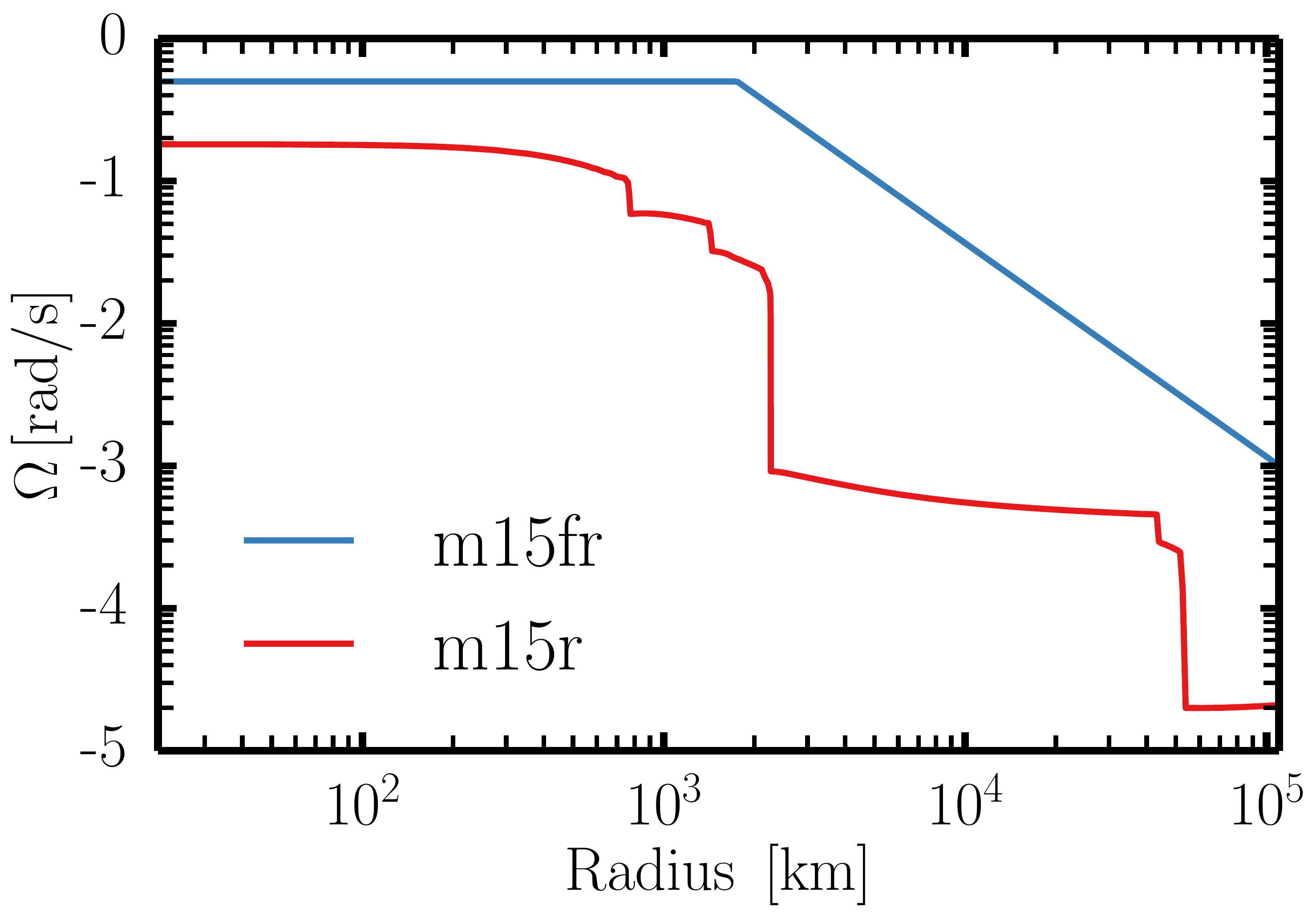}
\caption{Pre-collapse rotation profiles (\citet{Summa_18} assumed the angular
velocity to be constant on spheres) for
  the two rotating models m15fr and m15r, respectively. Both the x-axis and
  y-axis are shown in logarithmic scale.}
\label{figp2:rot}
\end{figure}

\citet{Summa_18} carried out five 3D simulations of the core-collapse
of the progenitor star with different initial rotation profiles and
different angular resolutions (see their Table 1 ). In two models of
different angular resolution the rotation profile dictated by the
stellar evolution calculations of \citet{heger_05} was used. In
another model \citet{Summa_18} used an arteficially enhanced rotation
rate and a modified rotation profile. Finally, they simulated two 3D
models of different angular resolution where the initial rotation rate
was set to zero throughout the star, \ie these two models were
non-rotating.  \fig{figp2:rot} shows the initial rotation profiles of
the rotating models.

For our study we selected the three models with the best angular
resolution, \ie models\ m15\_3D\_artrot\_2deg, m15\_3D\_rot\_2deg,
and m15\_3D\_norot\_4deg (see Table 1 in \citet{Summa_18}). In the
following we will use the shorthand model notation m15fr, m15r, and
m15nr, respectively, for these three models. The ending of the model names
indicates the rotation rate
( fr: \textbf{f}ast \textbf{r}otating , nr: \textbf{n}on \textbf{r}otating,
and r: slowly \textbf{r}otating ).

\begin{figure}         
\centering                            
\includegraphics[width=0.385\textwidth]{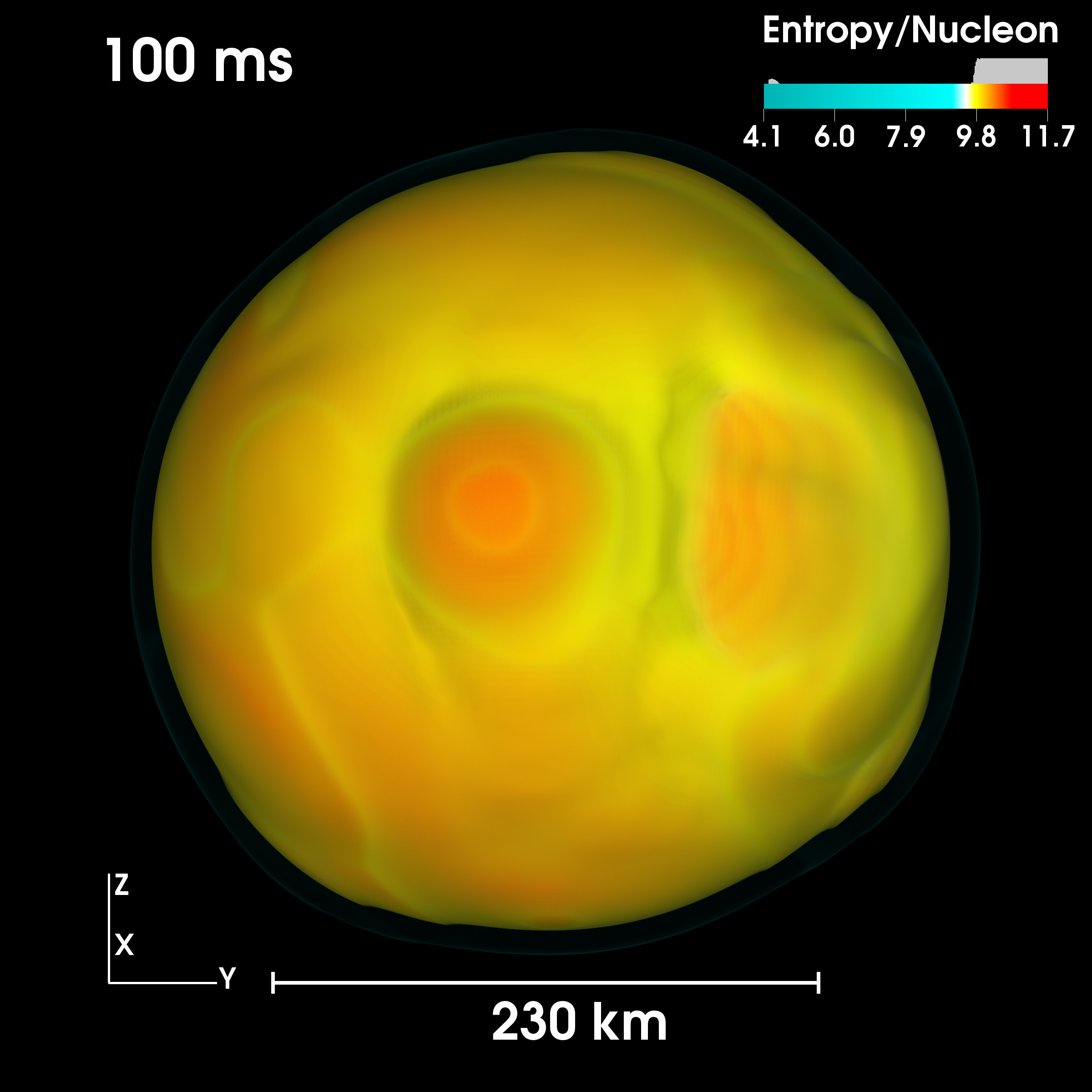} \\
\includegraphics[width=0.385\textwidth]{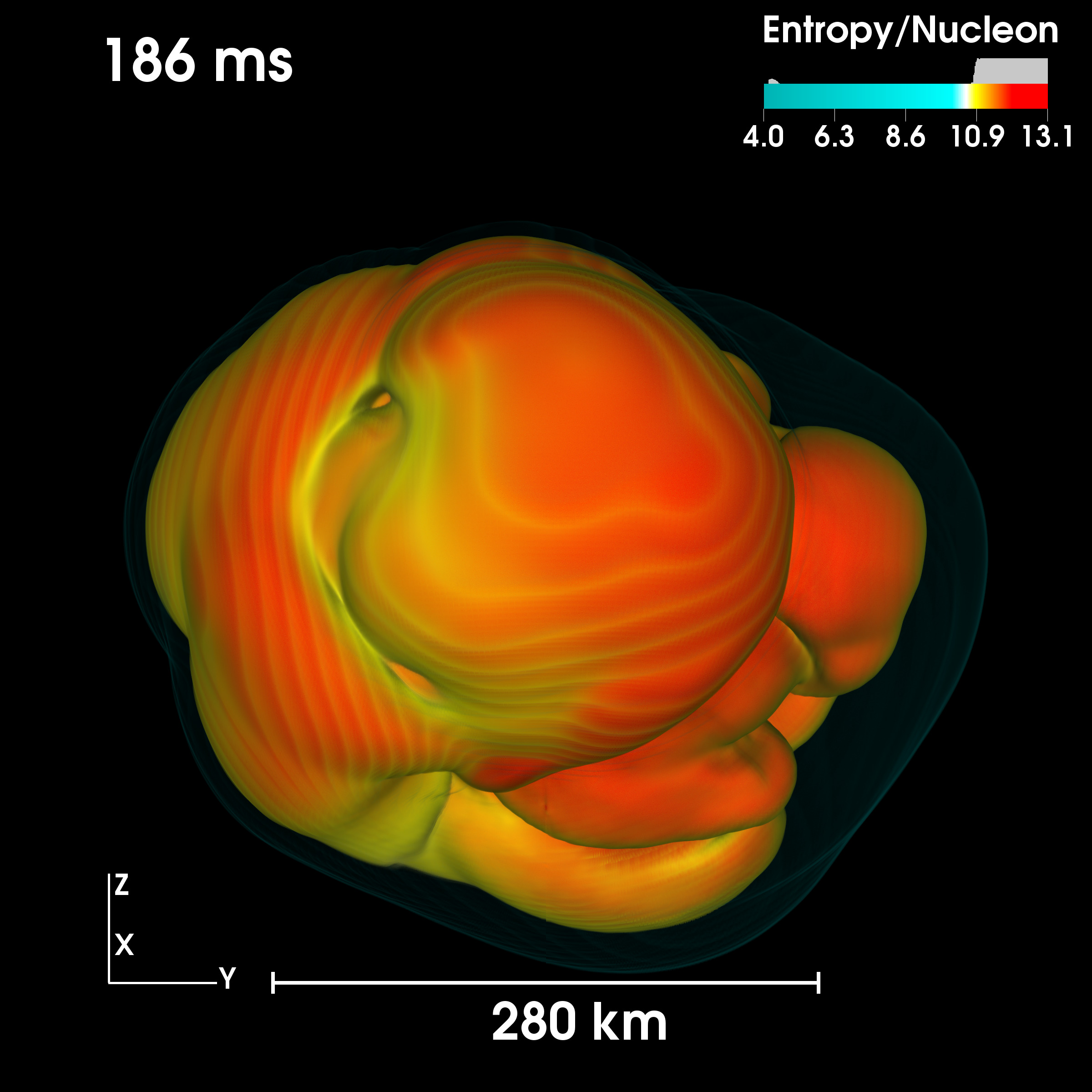} \\
\includegraphics[width=0.385\textwidth]{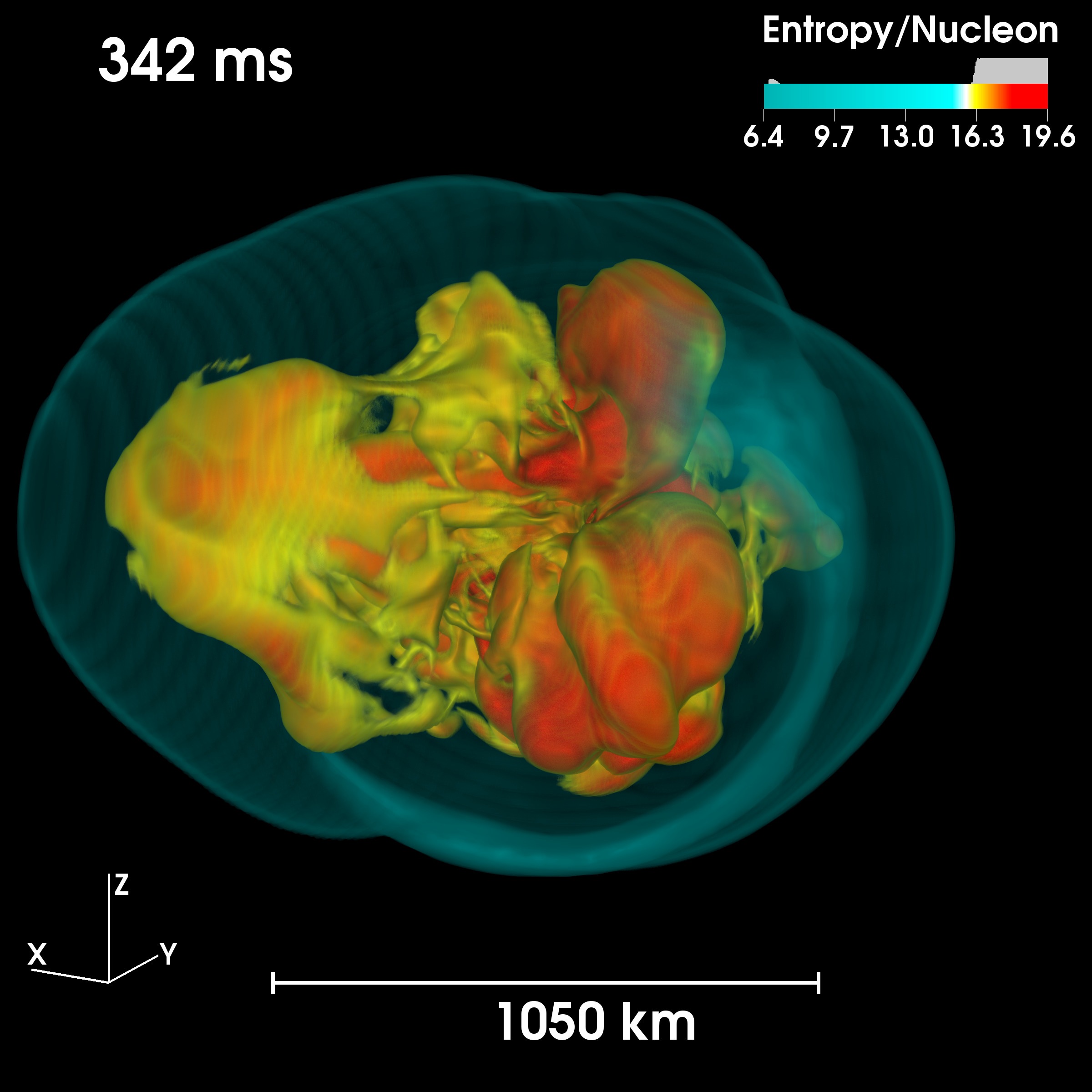}
\caption{Volume rendering of the entropy per nucleon for model m15fr
  at three different times, the time after core bounce is
  indicated in the upper left corner of each panel. The blue surface
  shows the shock front. The deformation of the shock front, which is
  indicative of SASI activity, can be seen in the middle panel. In
  the bottom panel the average shock radius has reached large values
  and runaway shock expansion is underway.}
\label{figp2:3dpics}
\end{figure}

\textbf{Model m15fr:} The initial rotation profile of model
  m15fr was set to a constant rotation rate of 0.5 rad/s throughout
  the inner 1731\,km of the core, while beyond this radius the
  rotation rate declines linearly (\fig{figp2:rot}).  The model was
  simulated using the Yin-Yang grid, the two grid patches had an initial
  resolution of 400, 56, and 144 zones in radial, polar, and azimuthal
  direction, respectively. This corresponds to a 2-degree angular
  resolution. The number of radial grid cell was increased during the simulation.
  After the initial shock expansion halts, around $60\,$ms
  post bounce, the average shock radius decreases slightly.  Between
  $\sim 80 - 160\,$ms post bounce the shock front is more or less
  stationary.  The shock starts to expand again at about $160\,$ms
  after core bounce and soon afterwards shock revival sets in.  Before
  shock revival the post-shock flow is dominated by a strong spiral
  SASI mode. The SASI sets in at around $\sim 100\,$ms post bounce. In
  \fig{figp2:3dpics} we show volume renderings of the entropy per
  nucleon, which gives an impression of the flow patterns of this
  model.

\begin{figure}         
\centering                            
\includegraphics[width=0.385\textwidth]{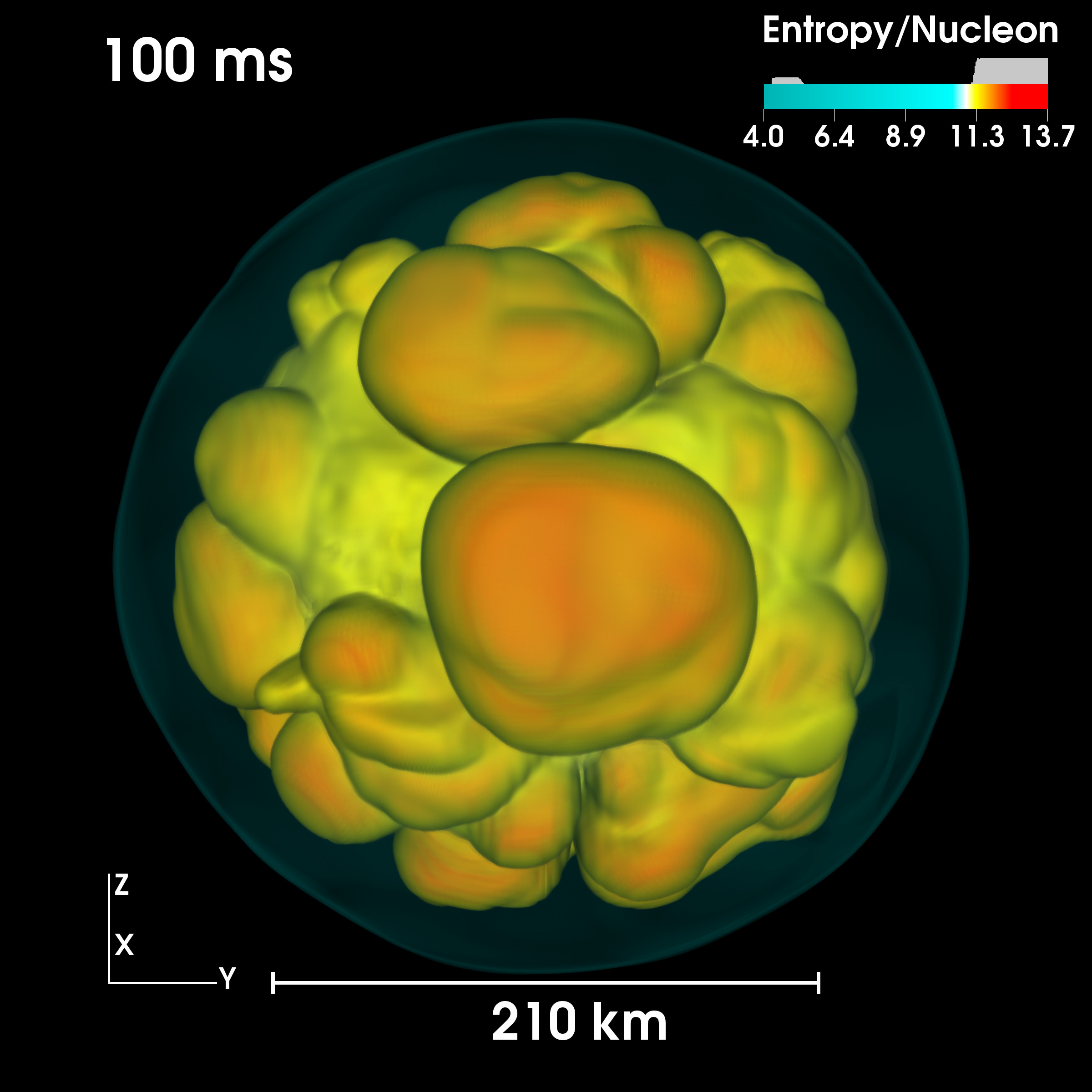} \\
\includegraphics[width=0.385\textwidth]{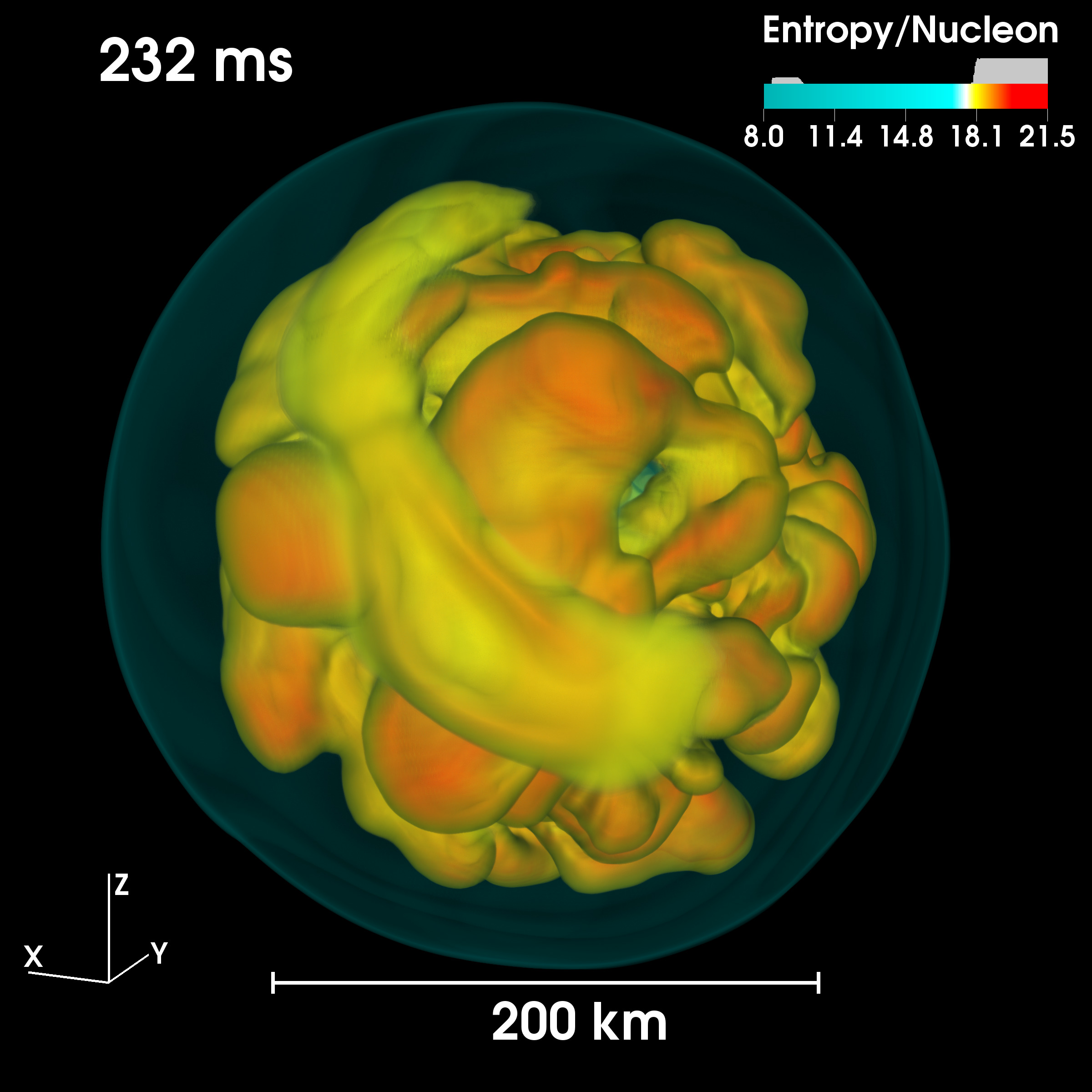} \\
\includegraphics[width=0.385\textwidth]{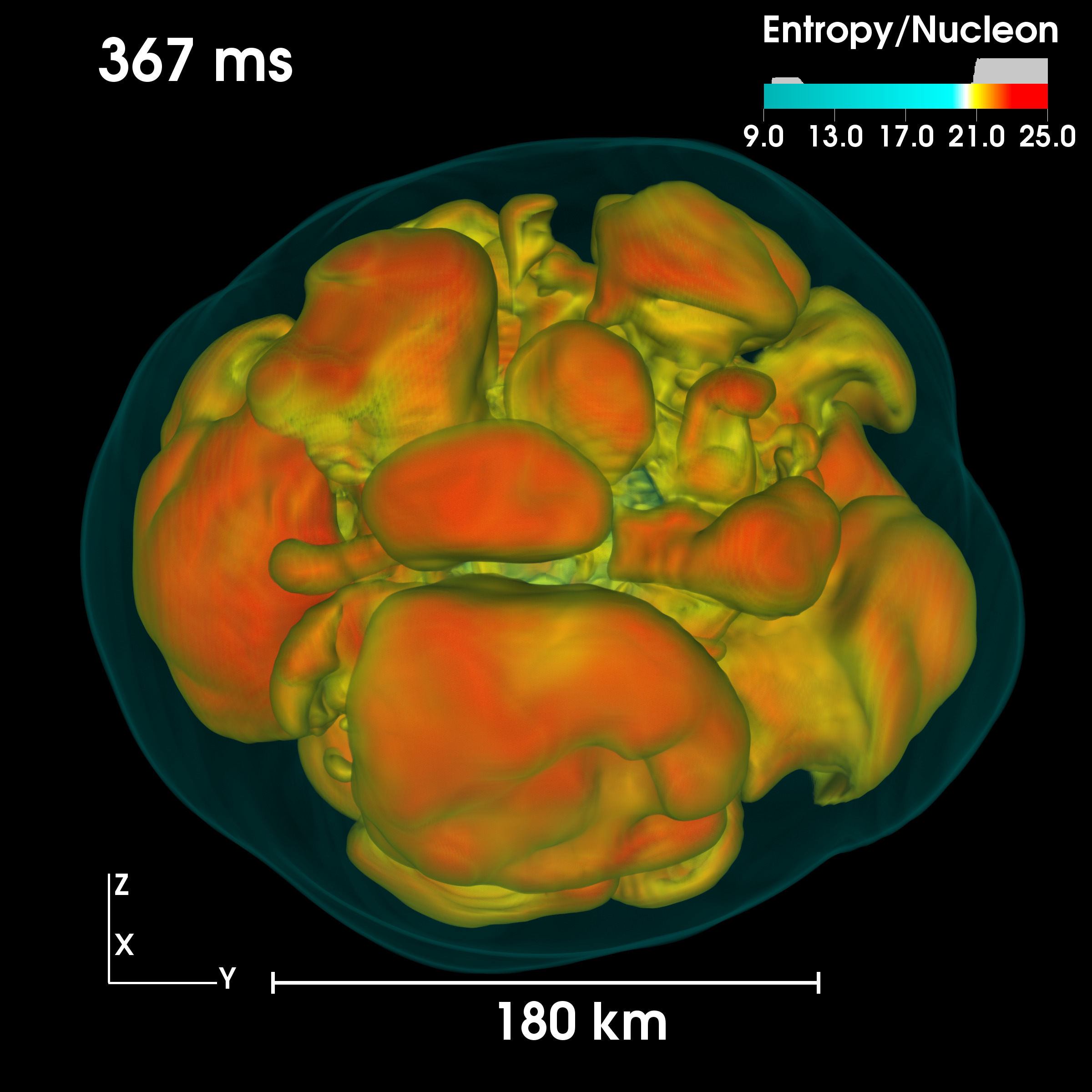}
\caption{Volume rendering of the entropy per nucleon for model m15r at
  three different times, the time after core bounce is indicated
  in the upper left corner of each panel.  The blue surface shows the
  shock front. In all panels, the flow in the region between the shock
  and the PNS is dominated by neutrino-driven convection, the typical
  mushroom shaped convective bubbles being clearly visible.  However,
  in the middle panel the convective activity is severely reduced. The
  yellow arc seen in this panel is a sign of weak SASI activity, which
  intermittently develops in model m15r.}
\label{figp2:3dpics2}
\end{figure}

  \textbf{Model m15r:} The initial rotation profile of model m15r
  (see \fig{figp2:rot}) was exactly that of the progenitor of
  \citep{heger_05}.  The model was simulated using the Yin-Yang grid,
  the two grid patches had an initial resolution of 400, 56, and 144 zones in
  the radial, polar, and azimuthal direction, respectively. This
  corresponds to a 2-degree angular resolution.
  The number of radial grid cell was increased during the simulation.
  For the first $\sim 80\,$ms after bounce the evolution of the average shock radius
  closely resembles that of model m15fr. However, around $100\,$ms
  after bounce the average shock radius starts to decrease.
  Only very weak, compared to the other two models, SASI oscillations
  develop in this model.
  The flow in the post-shock region is instead dominated by
  hot-bubble convection (see \fig{figp2:3dpics2}). However, between
  $\sim$ 200 and 260\,ms after bounce there is a period of low-amplitude
  dipole deformation of the shock front (see \fig{figp2:sasi}).  The
  decrease of the average shock radius, which started around 100\,ms
  after bounce, continues until $\sim 200\,$ms post bounce when the
  Si-O shell interface falls through the shock.  The decreased density
  ahead of the shock reduces the ram pressure and a transient period
  of shock expansion occurs. About $240\,$ms post bounce the
  expansion subsides and the shock front once more begins to retreat,
  a trend which continues until the end of the simulation.

  \textbf{ Model m15nr:} In model m15nr the initial rotation rate
  was set to zero at all radii. The model was simulated using the
  Yin-Yang grid, the two grid patches had an initial resolution of
  400, 28, and
  72 zones in radial, polar, and azimuthal direction,
  respectively. This means that model m15nr was simulated with an
  angular resolution that is two times coarser than the other two
  models (4 degrees instead of 2 degrees).
  The number of radial grid cell was increased during the simulation.
  The lower angular resolution discourages neutrino-driven convection and is
  conducive to the development of SASI activity. Initially, the shock
  expands and reaches a local maximum around $60\,$ms after bounce.
  Later on the shock radius steadily decreases until the Si-O shell
  interface falls through the shock. The decreased accretion rate
  leads to a transient period of shock expansion until the shock
  eventually starts to recede once again. Except for the fact that the
  average shock radius at late times is generally larger in model m15nr than in
  model m15r, model m15nr behaves similarly to model m15r in terms of
  the evolution of the average shock radius. However, unlike model
  m15r, model m15nr develops strong SASI activity, which is dominated
  by the spiral mode. The mode develops at about $120\,$ms after
  core bounce and peaks at $\sim 230\,$ms post bounce. Afterwards the
  SASI mode gradually decays towards the end of the simulation.
%\end{itemize}

\begin{figure}         
\centering                            
\includegraphics[width=0.45\textwidth]{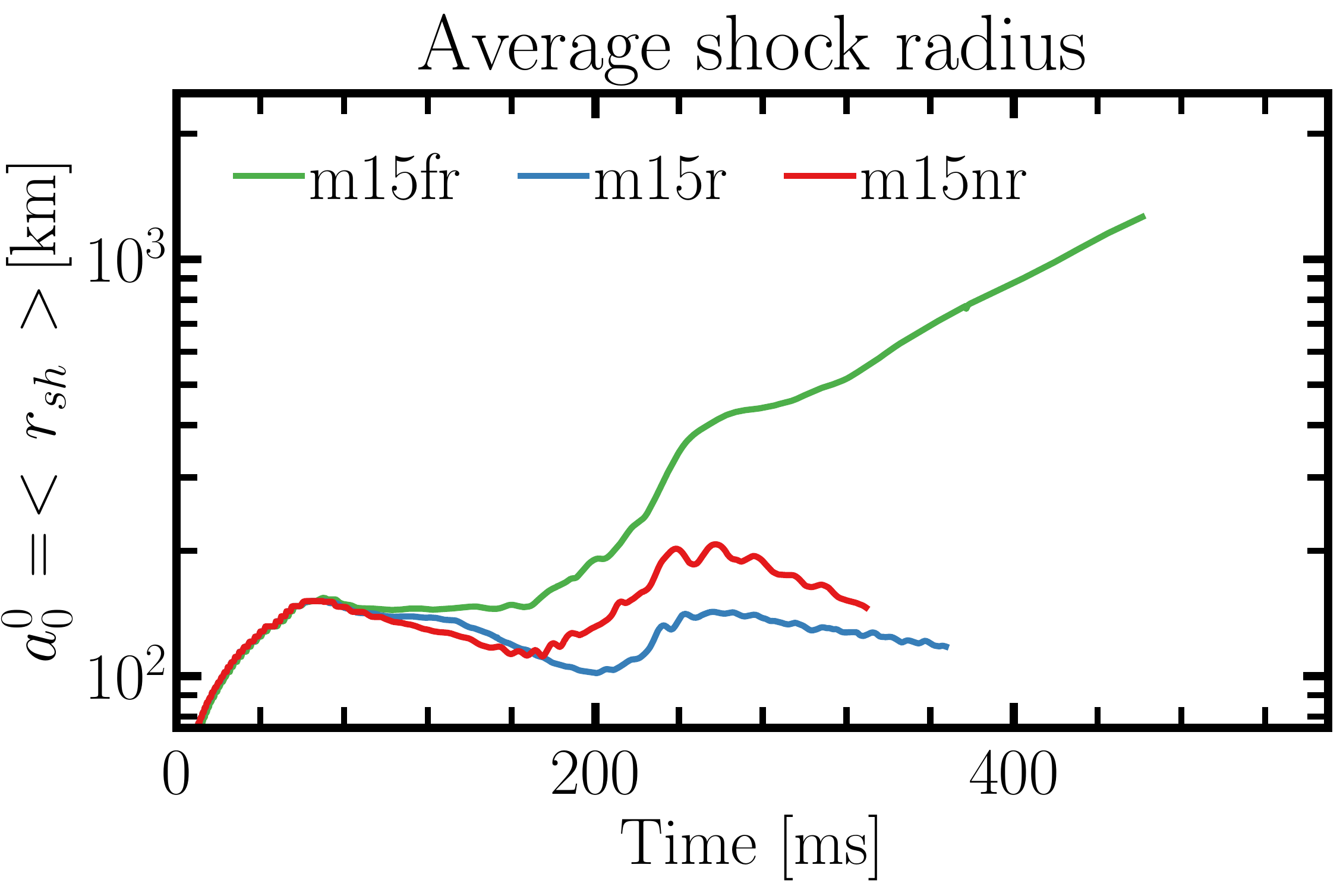}
\caption{Average shock radius for models m15fr (green), m15r (blue),
  and m15nr (red) as a function of time after core bounce. The average
  shock radius is defined as the $(l,m) = (0,0)$ expansion coefficient
  of the shock surface into spherical harmonics (\eq{eq:alsph}).}
\label{figp2:rsh}
\end{figure}

\begin{figure}         
\centering                            
\includegraphics[width=0.45\textwidth]{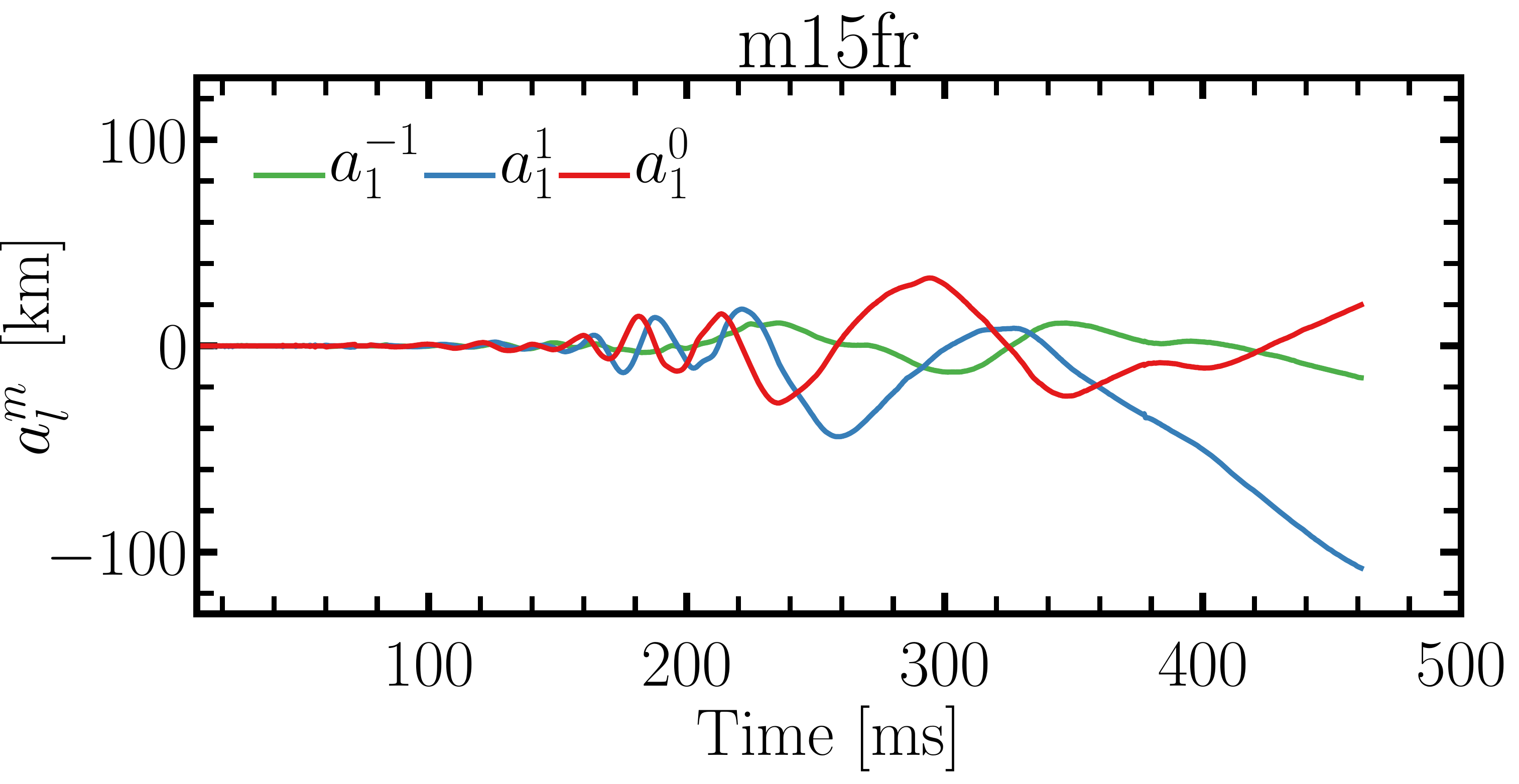}
\includegraphics[width=0.45\textwidth]{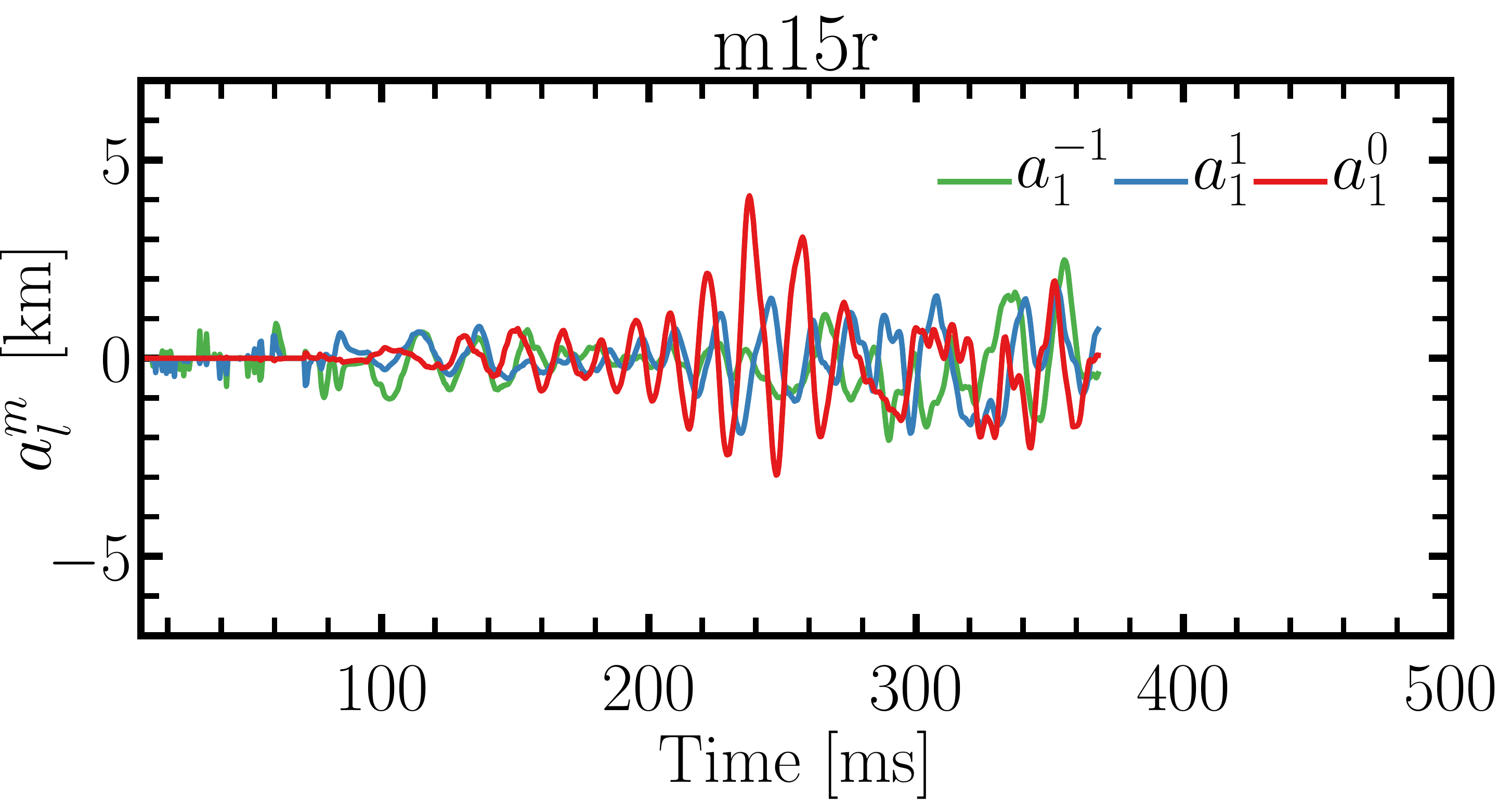}
\includegraphics[width=0.45\textwidth]{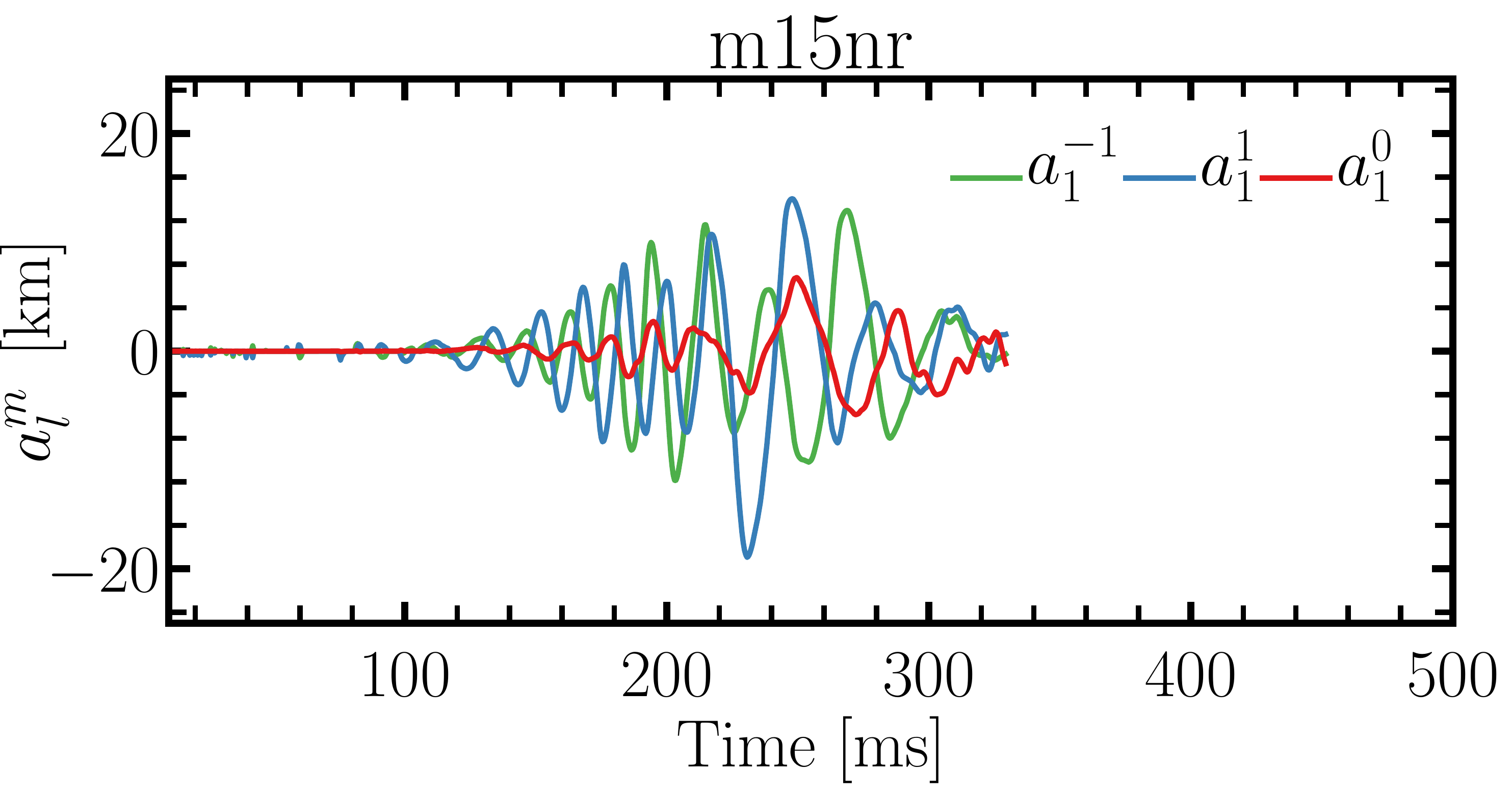}
\caption{The $(l,m) = (1,0)$, $(1,1)$, and $(1,-1)$ coefficients of,  the decomposition of the shock surface into spherical harmonics as a
  function of time after bounce (see \eq{eq:alsph}). From top to
  bottom: model m15fr, model m15r, and model m15nr.}
\label{figp2:sasi}

\end{figure}

Note that the coefficients given in \fig{figp2:rsh} and
\fig{figp2:sasi} are defined as 
\begin{equation} \label{eq:alsph} a_l^m(t_n) =
  \frac{(-1)^{|m|}}{\sqrt{4\pi(2l+1)}} \int
  r_{\mathrm{sh}}(\theta,\phi,t) Y^m_l \ud \Omega,
\end{equation}
where $r_{\mathrm{sh}}$ is the shock position %(given by the Riemann
%solver in the \textsc{Prometheus-Vertex} code)
, and $Y^m_l$ is the spherical harmonic of degree $l$ and order $m$.


%%%%%%%%%%%%%%%%%%%%%%%%%%%%%%%%%%%%%%%%%%%%%%%%%%%%%%%%%%%%%%%%%%%%%%%%%%%%
\section{Gravitational Wave Extraction} 
\label{sec:gw} 
The GW signals are extracted from the hydrodynamical simulations by
post-processing the output data using the quadrupole stress formula
\citep{finn_89,nakamura_89,blanchet_90}.  The formalism is described
in detail in, for example, \cite{finn_89}. Here we will give only the final formula.

The expressions for the two independent components, $h_{+}$ and
$h_{\times}$, of the gravitational wave tensor at observer distance $D$
in the transverse traceless (TT) gauge for a wave propagating into a general direction
given by the spherical polar coordinates $(r, \, \theta, \, \phi)$ are
\begin{align}
\label{eqT:hp}
  h_{+} = \frac{G}{c^4 D} & \Big[ \ddot{Q}_{11} (\cos^2{\phi} -
                            \sin^2{\phi} \cos^2{\theta})  \\ 
\nonumber
                         & + \ddot{Q}_{22} (\sin^2{\phi} -
                           \cos^2{\phi} \cos^2{\theta}) -
                           \ddot{Q}_{33} \sin^2{\theta} \\ 
\nonumber
                         & - \ddot{Q}_{12} (1 + \cos^2{\theta}) +
                           \ddot{Q}_{13} \sin{\phi} \sin{2\theta} \\ 
\nonumber
                         & + \ddot{Q}_{23} \cos{\phi} \sin{2\theta} \Big]
\end{align}
and
\begin{align}
\label{eqT:hc}
  h_{\times} = \frac{G}{c^4 D} & \Big[ (\ddot{Q}_{11} - \ddot{Q}_{22})
                                 \sin{2\phi}\cos{\theta}\\ 
\nonumber
                             & +\ddot{Q}_{12} \cos{\theta} \cos{2\phi}
                               - \ddot{Q}_{13} \cos{\phi} \sin{\theta}
  \\ 
\nonumber
                             & +  2\ddot{Q}_{23} \sin{\phi} \sin{\theta} \Big]. 
\end{align}
Here
\begin{equation} 
\label{eq:STFQ} 
\ddot{Q}_{ij} = \mathrm{STF} \left [2 \int \ud^3 x \, \rho
                \left ( v_i v_j - x_i \partial_j \Phi \right) \right].
\end{equation}
are the second-order time derivatives of the Cartesian components of
the quadrupole moment tensor (with $i,j = 1,2,3$) in the TT-gauge, which
are given in a form where the time derivatives have been elimated for
numerical reason by using the continuity and momentum equations
\citep{oohara_97, finn_89, blanchet_90}.

The other quantities in Eq.\ref{eq:STFQ} are the Cartesian velocity
components $v_i$, the Cartesian coordinates $x_i$, and the
gravitational potential $\Phi$ (including post-Newtonian corrections
used in the simulations).  $\mathrm{STF}$ denotes the projection
operator onto the symmetric trace-free part.

In the following, we give the GW signal strength not in terms of $h_+$
and $h_{\times}$, but in terms of \emph{GW amplitudes} which are
defined as
\begin{align} 
\label{eqT:zhchx}
A_{+} \equiv  D h_{+}, \qquad A_{\times} &\equiv D h_{\times}.
\end{align}
These GW amplitudes are convenient because they do not require us to
specify the distance $D$ between the observer and the source of the GWs.

Under the assumption of axisymmetry there is only one independent GW
component
\begin{equation} 
\label{eqT:htht} 
 \bmath{h}^\mathrm{TT}_{\theta \theta} = 
    \frac{1}{8}\sqrt{\frac{15}{\pi}} \sin^2{\vartheta} \frac{A_{20}^\mathrm{E2}}{D},
\end{equation}
where $\vartheta$ is the inclination angle of the observer with
respect to the axis of symmetry, and $A_{20}^\mathrm{E2}$ is the only
non-zero component of the quadrupole moment.  In spherical coordinates
$A_{20}^\mathrm{E2}$ is given by
\begin{eqnarray} \label{eq:2dquad}
  A_{20}^\mathrm{E2} (t) =  \frac{G}{c^4} \frac{16 \pi^{3/2}}{\sqrt{15}} 
  \int_{-1}^{1}\int^{\infty}_0 \rho \left [ v_r^2(3 \zeta^2 - 1)+ \right. \nonumber \\
  v_{\theta}^2(2-3 \zeta^2) - v_{\phi}^2 - 6 v_r v_{\theta} \zeta \sqrt{1-\zeta^2}  \nonumber \\
  -(3 \zeta^2 - 1) r \partial_r \Phi  +\left. 3 \zeta
  \sqrt{1-\zeta^2} \partial_{\theta} \Phi \right ]r^2 dr \, d\zeta,
\end{eqnarray}
where, $\partial_i$ ($i = r, \theta, \phi$) and $v_i$ are the
derivatives and velocity components, respectively, along the basis
vectors of the spherical coordinate system, and $\zeta$ is a short
hand notation for $\cos \theta$.  See, e.g., \cite{mueller_97} for
details.

%%%%%%%%%%%%%%%%%%%%%%%%%%%%%%%%%%%%%%%%%%%%%%%%%%%%%%%%%%%%%%%%%%%%%%%%%%%%
\section{Results} \label{sec:res}
\subsection{Qualitative description of the gravitational wave signals}
Since the 3D simulations of \citet{Summa_18} do not cover the phase of
collapse and bounce, the GW signal emitted during the earlier
evolution ($t \la 10\,$ms post-bounce) was obtained from their
preceding 2D simulations. This part of the GW signal we will discuss
later in section ~\ref{sec:cb}.

\begin{figure*}
\centering                            
\includegraphics[width=0.99\textwidth]{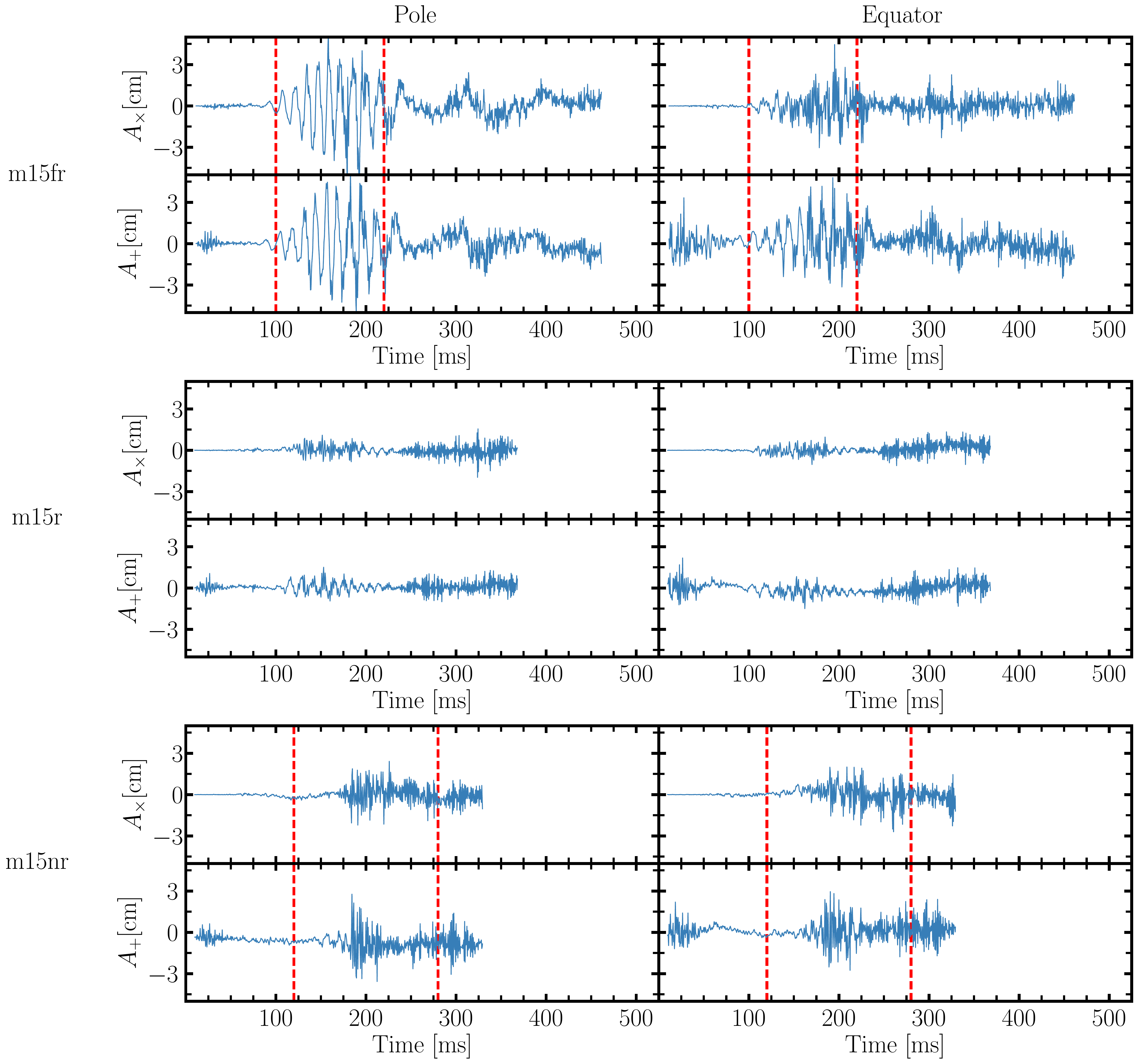}
\caption{GW amplitudes $A_+$ and $A_\times$ as functions of time after
  core bounce for models m15fr, m15r, and m15nr (from top to bottom).
  The two columns show the amplitudes for an observer situated along
  the pole (left) and in the equatorial plane (right).  Episodes of
  strong SASI activity occur between the vertical red dashed lines.}
\label{figp2:amps}
\end{figure*}

\begin{figure*}
\centering                            
\includegraphics[width=0.9\textwidth]{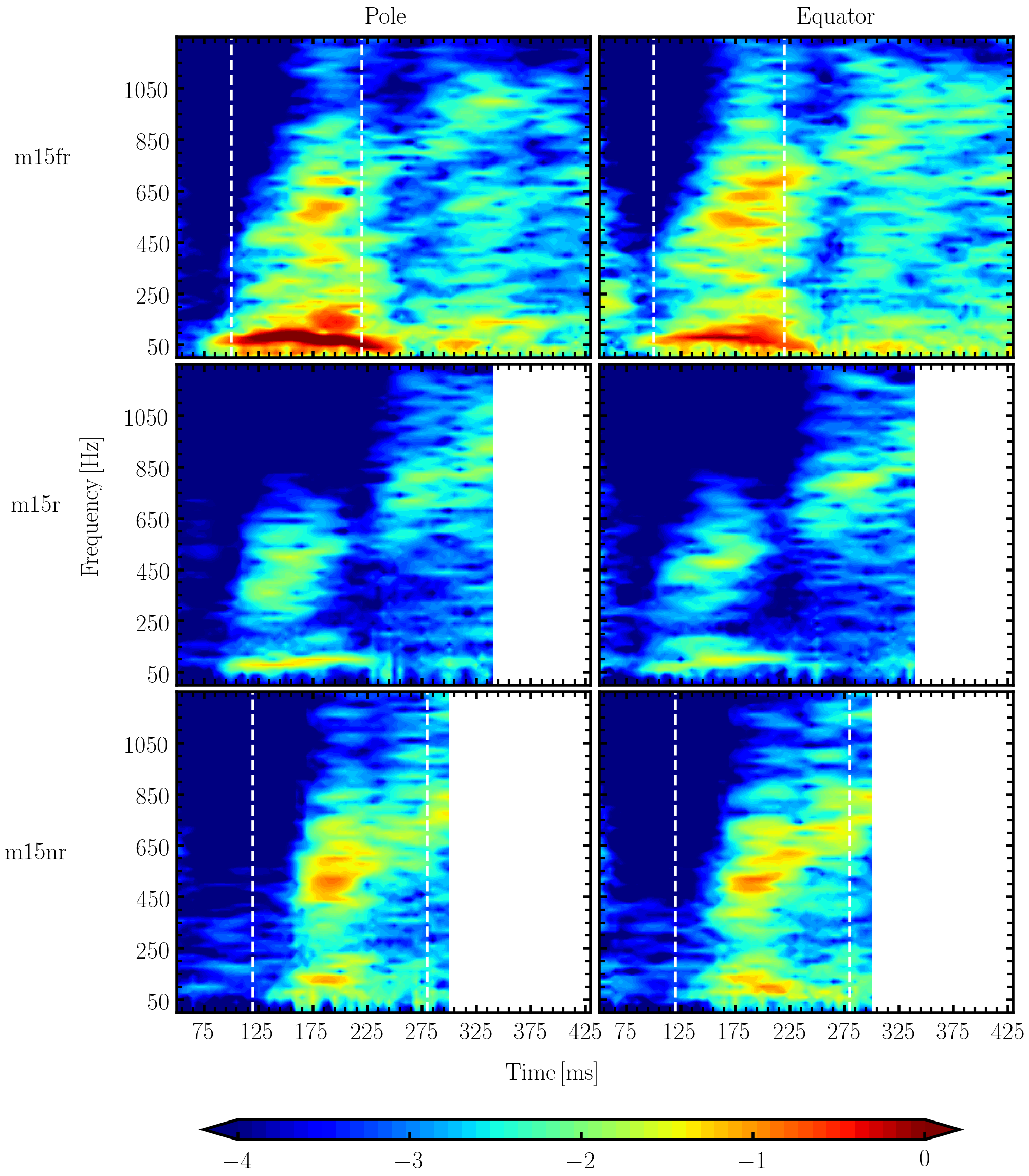}
\caption{Amplitude spectrograms for a sliding window of 50\,ms width
  summed over the two polarisation modes,\ie
  ($|\text{STFT}[{A_+}]|^2 + |\text{STFT}[{A_{\times}}]|^2$). The rows
  show the results for models m15fr, m15fr, and m15nr (from top to
  bottom), and the two columns give the amplitudes for an observer
  situated along the pole (left) and in the equatorial plane (right),
  respectively. Time is measured in ms after core bounce. Vertical
  dashed lines bracket SASI episodes. The amplitudes are normalised by
  the same factor, and their values are colour-coded on a logarithmic
  scale.}
\label{figp2:spec}
\end{figure*}

\fig{figp2:amps} shows the GW amplitudes generated by asymmetric mass
motions for two different observer orientations for the three models
of \citet{Summa_18} that we have analyzed.  The two columns represent
observers located along the pole (left) and in the equatorial plane
(right), respectively, of models m15fr, m15r, and m15nr (from top to
bottom), \ie in order of decreasing initial rotation rate. Vertical
red dashed lines mark beginning and end of episodes of strong SASI
activity.  The corresponding amplitude spectrograms are shown in
Fig.~\ref{figp2:spec}, which we obtained by applying the short-time
Fourier transform (STFT) to the corresponding waveforms. To calculate
the STFT we moved a window of 50\,ms width across the discrete GW
wave amplitude data applying the discrete Fourier transform (DFT) to
each window.  We define the DFT as
\begin{equation} \label{eq:DFT}
\widetilde{X}_k (f_k) = \frac{1}{M}  \sum^M_{m=1} x_m e^{-2\pi i k m/N},
\end{equation}
where $x_m$ is the time series obtained by sampling the underlying
continuous signal at $M$ equidistant discrete times, and $f_k = k/T$
is the frequency of bin $k$ with $T$ being the total duration of the
analyzed GW signal (50 ms in the considered case).
The amplitude spectrograms in \fig{figp2:spec}
show the square of the STFT amplitudes summed over of the cross and
plus polarisation modes, \ie
$|\text{STFT}[{A_+}]|^2 + |\text{STFT}[{A_{\times}}]|^2$.  We
convolved the GW amplitudes with a Kaiser window (shape parameter
$\beta = 2.5$) before we applied the DFT, whereby frequencies below
$50\,$Hz and above $1100\,$Hz are filtered out.

We find that the GW signals of the two rotating models m15fr and m15r
do not differ fundamentally from the signals of the non-rotating model
m15nr and those models discussed in \cite{andresen_17}.  In all three
models considered here, an initial phase of quiescence is followed by
a phase of very active, but case-dependent, emission during
which the amplitudes are of the order of a few centimeters.
As expected, the fast rotating model m15fr
emits the strongest GW signal, while unexpectedly the non-rotating
model m15nr shows larger amplitudes than the moderately rotating model
m15r.  Hence, we find no clear correlation between the initial
rotation rate and the strength of the GW emission, which might point
to considerable stochastic variation. The spectrograms
(\fig{figp2:spec}) of all three models show the low-frequency
($\nu_\mathrm{GW} \la 250\,$Hz) and high-frequency
($\nu_\mathrm{GW} \ga 250\,$Hz) signal components reported by
\cite{kuroda_16} and \cite{andresen_17}.

{Note
that the signal associated with prompt-convection is present in all three
models, but can only be seen in $A_+$. We suspect that this unexpected behavior is a
consequence of an erroneous treatment of fictitious forces in the 2D simulations
that evolved the core through bounce \citep{Summa_18}. For this reason
angular momentum was not strictly conserved during collapse. This led to
slower rotation rates in the inner layers of the core. When the terms switched off by mistake during
collapse were taking into account in 3D after core bounce, transient oscillations (caused by the
adjustment of the models) affected the first $\sim 100\,$ms
after bounce and are the reason why the signals exhibit such peculiar GW emission at
early times. The signals during this phase should
look more like the ones presented in \cite{andresen_17}.}

\textbf{Model m15fr:} The GW signal from model m15fr is characterised
by strong emission over a broad range of frequencies. The pronounced
low-frequency and high-frequency emission mentioned above is clearly
visible (see \fig{figp2:spec}, upper row), but covering a broader
frequency range than in the other two models. At about 200\,ms
post-bounce the two emission regions almost overlap. When
run-away shock expansion is fully underway at $\approx 250\,$ms post-bounce),
the overall GW amplitudes strongly decrease, but both low-frequency
and high-frequency emission continue to be present until the end of
the simulation. Coinciding with the onset of shock expansion, the central
SASI frequency starts to decrease around $\approx 200\,$ms after bounce.

\textbf{Model m15r:} With GW amplitudes never exceeding
$1.5\,$cm, this model produces the weakest GW signal of the three models.
Furthermore, the signal is strongly reduced in the time
period between $180$ and $250\,$ms post-bounce, when the
high-frequency emission almost completely subsides and only a very
weak low-frequency emission is recognizable.  At $\sim 250\,$ms
post-bounce the high-frequency emission starts to increase again,
while the emission at low-frequencies almost ceases.

\textbf{Model m15nr:} The non-rotating model is characterized by a
relatively long initial quiescent phase compared to the two rotating
models. This could be a consequence of the lower
angular resolution (4 degrees instead of 2 degrees ) of this simulation,
which delays the growth of convection in the post-shock region.
Weak low-frequency GW emission sets in $\sim 125\,$ms after
bounce. This low-frequency signal component increases in strength
until it reaches a maximum value at $\sim 175\,$ms post-bounce.
Approximately $25\,$ms after the onset of low-frequency emission,
high-frequency emission develops at $\sim 150\,$ms post-bounce. Both
signal components remain present until the end of the simulation,
however, varying considerably in strength.

\subsubsection{Time-integrated energy spectra}
\begin{figure}
\centering                            
\includegraphics[width=0.95\linewidth]{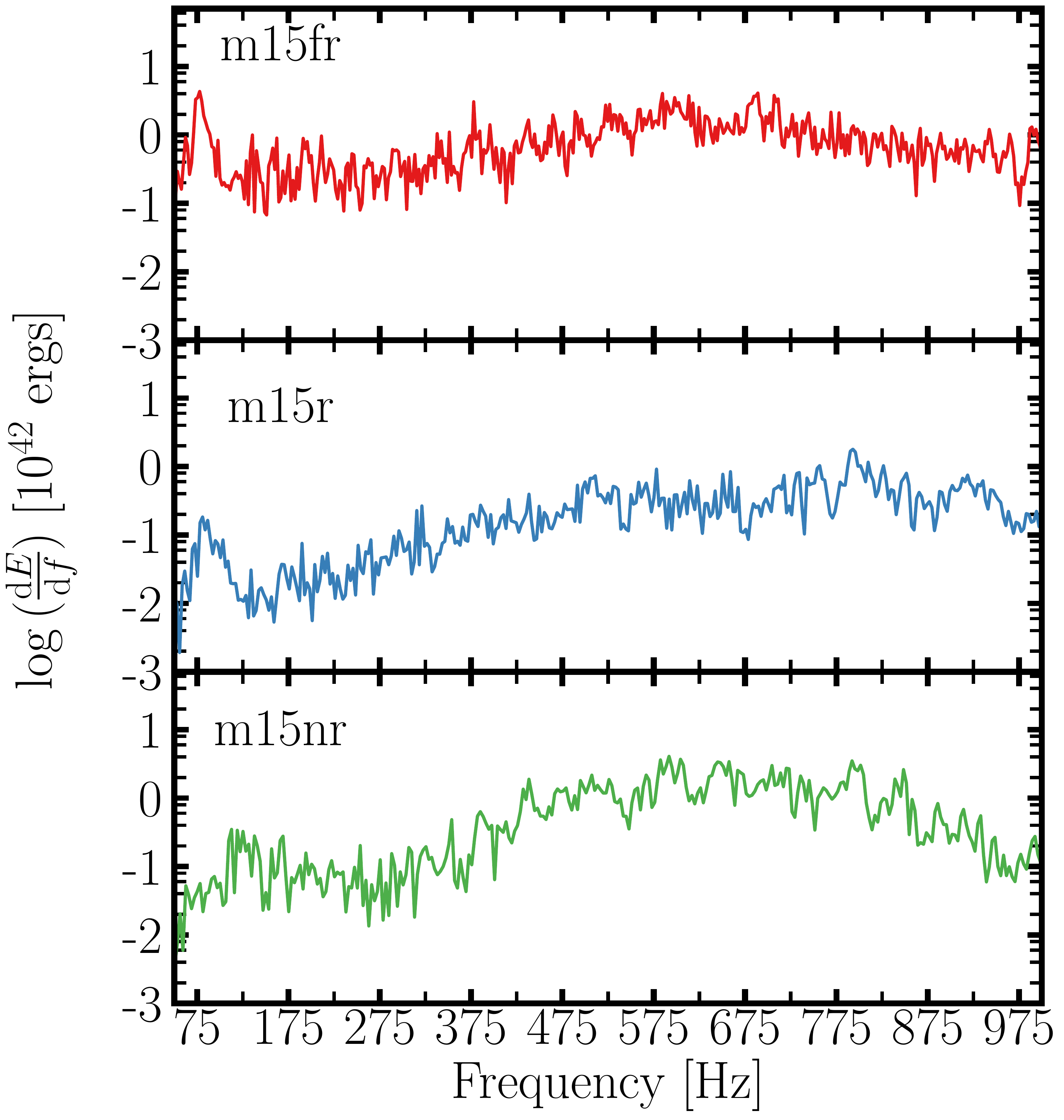}
\caption{Time-integrated GW energy spectra $\ud E/ \ud f$ for models
  m15fr, m15r, and m15nr (top to bottom). The spectra are computed
  for the full time duration of the simulations. }
\label{figp2:energy_spectra}
\end{figure}

Time-integrated energy spectra, $\mathrm{d} E / \mathrm{d} f$, for
each of the models are shown in Fig.~\ref{figp2:energy_spectra}. These
spectra are calculated from the Fourier transform of the second
time-derivatives of the Cartesian components of the mass quadrupole
moments for every bin $k$ according to
\begin{equation} \label{eq:enspc} 
  \frac{\mathrm{d E}}{\mathrm{d} f} 
  \approx  \Bigg[\frac{\Delta E}{\Delta f}\Bigg]_k 
  = \frac{2G}{5 c^5} (2\pi f_k)^2 \big|\widetilde{\ddot{Q}^{ij}}_k
    \widetilde{\ddot{Q}^{ij}}_k| \, T^2.
\end{equation}
Here $T$ represents the duration of the simulations.
See \cite{andresen_phd} for details. 

The time-integrated energy spectrum of model m15fr is rather flat,
with strong emission over a wide range of frequencies and a local
maximum at $\sim 75-100\,$Hz. The slower rotating model m15r, which
does not develop strong SASI oscillations, emits much less energy at
low frequencies, but also exhibits a maximum at $\sim 75-100\,$Hz.
The maximum, however, is far less pronounced than in model m15fr.  The
energy spectrum of model m15nr is a hybrid of the spectra of the two
other models.  As for model m15fr, there is a significant amount of
energy radiated away by GW below 300\,Hz, but the spectrum is less
flat than the one of model m15fr.

%%%%%%%%%%%%%%%%%%%%%%%%%%%%%%%%%%%%%%%%%%%%%%%%%%%%%%%%%%%%%%%%%%%%%%%%%%%%
\section{Excitation of gravitational waves} 
\label{sec:p2ext}
\cite{andresen_17} studied the GW signals from four 3D simulations of
core-collapse supernovae
\citep{tamborra_13,hanke_13,tamborra_14a,tamborra_14b,hanke_phd,melson_15b},
based on three \emph{non-rotating} progenitors. They found that strong
SASI activity excites low-frequency GW emission by creating an
asymmetric mass distribution in the post-shock layer which directly
leads to GW emission. In addition, the high-velocity violent downflows
resulting from SASI activity perturb the PNS surface and even the PNS
convection layer and excite non-resonant g-modes in the PNS, which in
turn lead to strong GW emission.  Both the frequency of the forced
g-modes in the PNS surface and the mass motions in the post-shock
layer are set by the typical time scale of the SASI oscillations as
pointed out by \citet{kuroda_16}.  Downflows from the post-shock layer
onto the PNS also excite resonant surface g-modes and so do convective
plumes from the PNS interior overshooting into the outer layer of the
PNS \citep{marek_09,murphy_09,mueller_13,morozova_18}.  The
propagation of these g-modes is responsible for the high-frequency
emission.  \cite{andresen_17} found that they are mainly excited by
PNS convection, with a small contribution from downflows impinging on
the PNS surface from above.
For the most part, these findings also hold for the three
  models presented here, but there are two notable exceptions. The GW
  signal generated by PNS convection ($f_{\rm GW} \ga 300\,\mathrm{Hz}$)
  is weak in the two rotating models (see Fig.\,7).
  Additionally, in model m15fr the low-frequency emission
  ($f_{\rm GW} \la 300\,\mathrm{Hz}$) generated by mass motions in the
  post-shock layer (see Fig.\,9) is stronger than what was
  found for the models of \cite{kuroda_16} and \cite{andresen_17}.

The importance of mass motions in the post-shock layer, with
  respect to the low-frequency signal, can be seen by dividing the
  simulation volume into three layers (A, B and C; see also
    Fig.\,4 in \cite{andresen_17}) and calculating the contributions
  from each layer to the integral in \eq{eq:STFQ}.  Layer~A consists
  of the inner PNS and contains the convectively active layer as well
  as the region where convective plumes overshoot into the
  convectively stable PNS surface layer.  The second layer, which we
  call layer~B, covers the region extending from the top of layer~A to
  the PNS surface. It should be noted that there are several ways to
  define the surface of the PNS and that we define it to be where the
  angle-averaged density drops below $10^{10}$ g/cm$^3$.  GWs excited
  by downflows onto the PNS should be emitted, at least in part, from
  layer~B.  Layer~C encompasses the volume between the PNS surface and
  the outer boundary of the simulation grid.  This outermost layer
  captures GWs produced by mass motions in the post-shock layer and by
  matter accreting onto the PNS surface. We refer the reader to
  \cite{andresen_17} for a more detailed definition of the three
  layers.

\begin{figure*}
\centering                            
\includegraphics[width=0.9\textwidth]{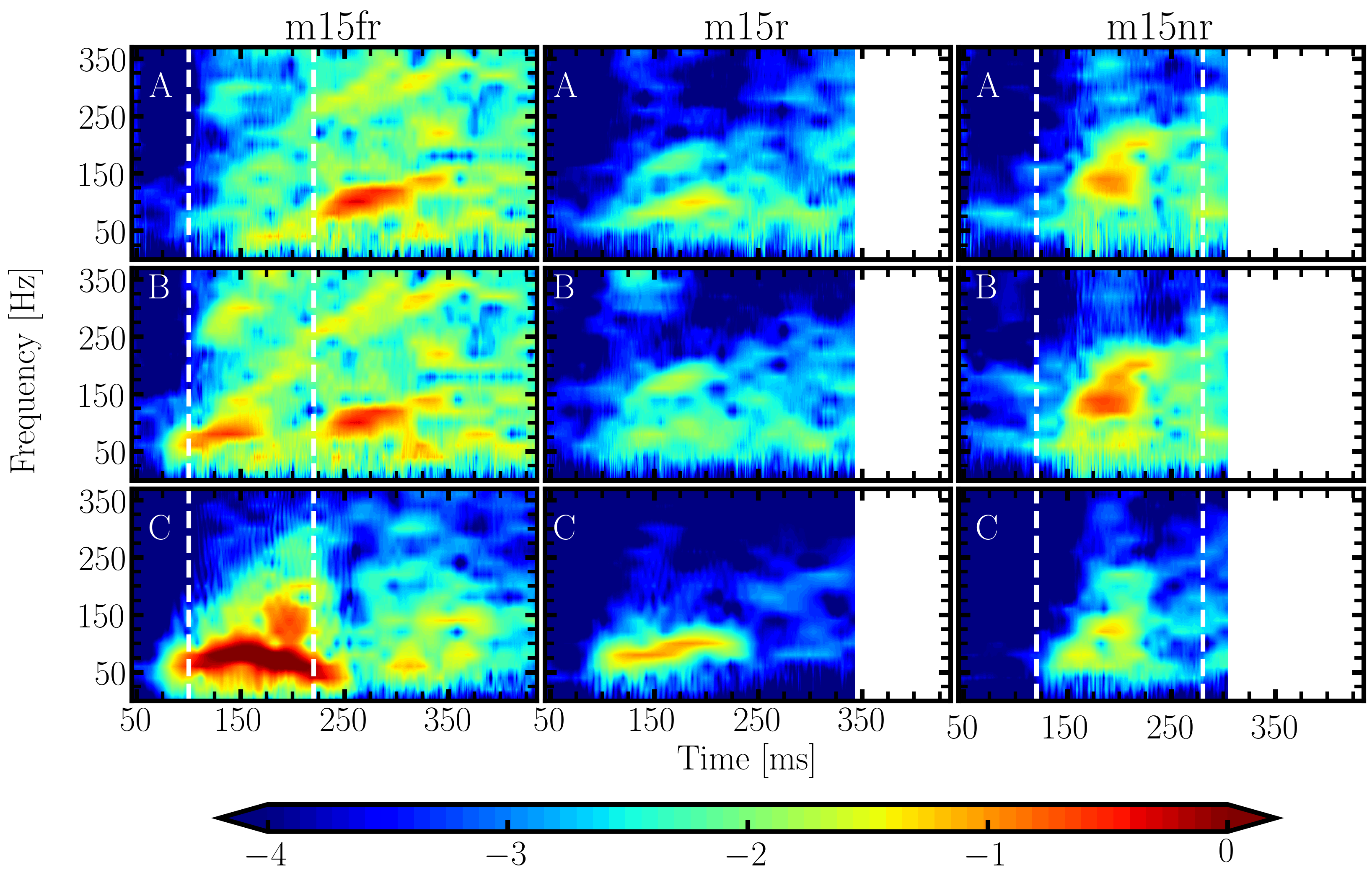}
\caption{Amplitude spectrograms for a sliding window of 50\,ms width
  summed over the two polarisation modes, ($|\text{STFT}[{A_+}]|^2 +
  |\text{STFT}[{A_{\times}}]|^2$).  The columns show the three
  different models m15fr, m15r, and m15nr (from left to right). The
  first row shows the contribution from layer~A, the middle row the
  contribution from layer~B, and the bottom row shows the contribution
  from layer~C. Time is measured in ms after core bounce. Vertical
  dashed lines bracket SASI episodes. The amplitudes are normalised by
  the same factor and their values are colour-coded on a logarithmic
  scale. We show the spectrograms for the polar observer direction.}
\label{figp2:cuts}
\end{figure*}

In \fig{figp2:cuts} we show the spectrograms of the GWs
  emitted in the three individual layers.  It should be emphasised
  that this plot has to be viewed with caution.  Contributions to
  \eq{eq:STFQ} from different layers that would normally cancel out
  can create artefacts.  In our simulations, matter is continuously
  flowing from one layer to another and these non-zero mass fluxes at
  the boundaries are problematic when calculating the GWs from
  separate layers.  The exact definition of the boundaries becomes
  important for the strength of individual features in the
  spectrograms. We have verified that the general picture stays the
  same when we shift our boundaries, within reasonable limits.

From \fig{figp2:cuts} we see that the low-frequency emission
  in model m15fr differs from that found by \cite{kuroda_16} and
  \cite{andresen_17}.  In model m15fr the strongest contribution to
  the low-frequency signal comes from layer~C, and not from the
  surface and internal layers of the PNS as reported by
  \cite{kuroda_16} and \cite{andresen_17}.  However,
  \cite{andresen_17} concluded that non-negligible contributions to
  the low-frequency emission arise from each of the three layers,
  which also holds for the models presented in this work.
  Because of its temporal coincidence with the SASI episode it is
  clear that the strong emission from layer~C seen in model m15fr
  is a consequence of the strong spiral SASI oscillations that develop
  soon after core bounce. The mass accretion rate decreases rapidly as
  a function of time after bounce (see Fig. 6 of \cite{Summa_18}). The
  large amount of high-density matter falling through the shock at
  early times combined with the strong oscillations of the shock
  should intuitively lead to strong gravitational wave emission
  produced by mass motions in the layer between the PNS surface and the shock.
  At later times, when the mass accretion rate drops significantly,
  or shock expansion sets in, less
  mass will be involved in mass motions induced by the SASI. Hence, it
  is reasonable to expect weaker GW emission.

While in the non-rotating model m15nr strong
    high-frequency emission is clearly visible throughout the whole
    simulation (see last row of \fig{figp2:spec}), we see a reduction
    in the emission above 250 Hz in the rotating model m15r, but the
    signal component is still clearly visible in the second row of
    \fig{figp2:spec}. In the fast rotating model m15fr strong
  high-frequency emission can be seen at early times, during the SASI
  phase, but once the SASI subsides the emission is drastically
  reduced. The emission at early times is caused by SASI-modulated
  downflows perturbing the PNS surface. A similar behavior was seen
  in model s20 and model s20s of \cite{andresen_17}. At late times,
  when the shock expansion is well underway, there is no SASI activity
  and the mass accretion rate onto the PNS is greatly reduced. In
  other words, the typical mechanisms that excite oscillations of the
  PNS from above are absent.  The lack of high-frequency emission in
  model m15fr, at late times (top row of \fig{figp2:spec}), indicates
  that PNS convection does not generate strong GW emission. The
  reduction of high-frequency emission from the PNS convective layer
  in the two rotating models is connected to the fact that the
  initial rotation leads to the development of a positive angular
  momentum gradient in the PNS convective layer.  As a result, the PNS
  convection is weakened according to the Solberg-Høiland criterion
  \citep{janka_01b,buras_06b}.
The basic physical picture is that when a buoyant plume propagates
outwards, with conserved specific angular momentum, its rotation rate
is less than that of the surrounding medium and it experiences a
weaker centrifugal force. The result is a net force which acts against
the outward propagation of the plume. Angular momentum transfer by
convection will eventually flatten the rotation profile within the PNS
and the restoring force will gradually decrease with time.
By considering the mean field kinetic energy equation, we can see that
the average kinetic energy contained in the convective region
decreases as we increase the initial rotation rate. It then follows
that less energy is injected into the overshooting region and
converted into g-modes, which in turn results in a weaker GW signal at
high frequencies.  In the mean field approach, where the flow is
decomposed into average and fluctuating terms, one finds that the
kinetic energy sources are buoyancy work due to density fluctuations
($W_b$) and work resulting from pressure fluctuations ($W_p$). Viscous
forces can also contribute to the overall kinetic energy budget, but
the fluid in our simulations is modelled as a non-viscous perfect
fluid.
If we only consider the radial direction, $W_b$ and $W_p$ can be written as
\begin{equation} \label{eq:Wb}                                                                        
  W_b(r) = \overline{\langle g \rho'\, v_r'\rangle},
\end{equation}                                                                                         
\begin{equation} \label{eq:Wp}
  W_p(r) = \overline{\langle p'\, \nabla_r  v_r'\rangle},
\end{equation}
where $\rho'$, $p'$, and $v_r'$ are the local deviations from the
angular averages of density, pressure, and radial velocity (at any
given radius), respectively. The gravitational acceleration is
represented by $g$ and $\nabla_r$ represents the radial part of the
divergence operator.  The angle brackets denote averaging over all
angular bins and the overbar represents time averaging
(\cite{hurlburt_86}, \cite{hurlburt_94}, \citet{nordlund_09}, and \cite{viallet_13}).
\begin{figure}      
  \centering
  \includegraphics[width=0.48\textwidth]{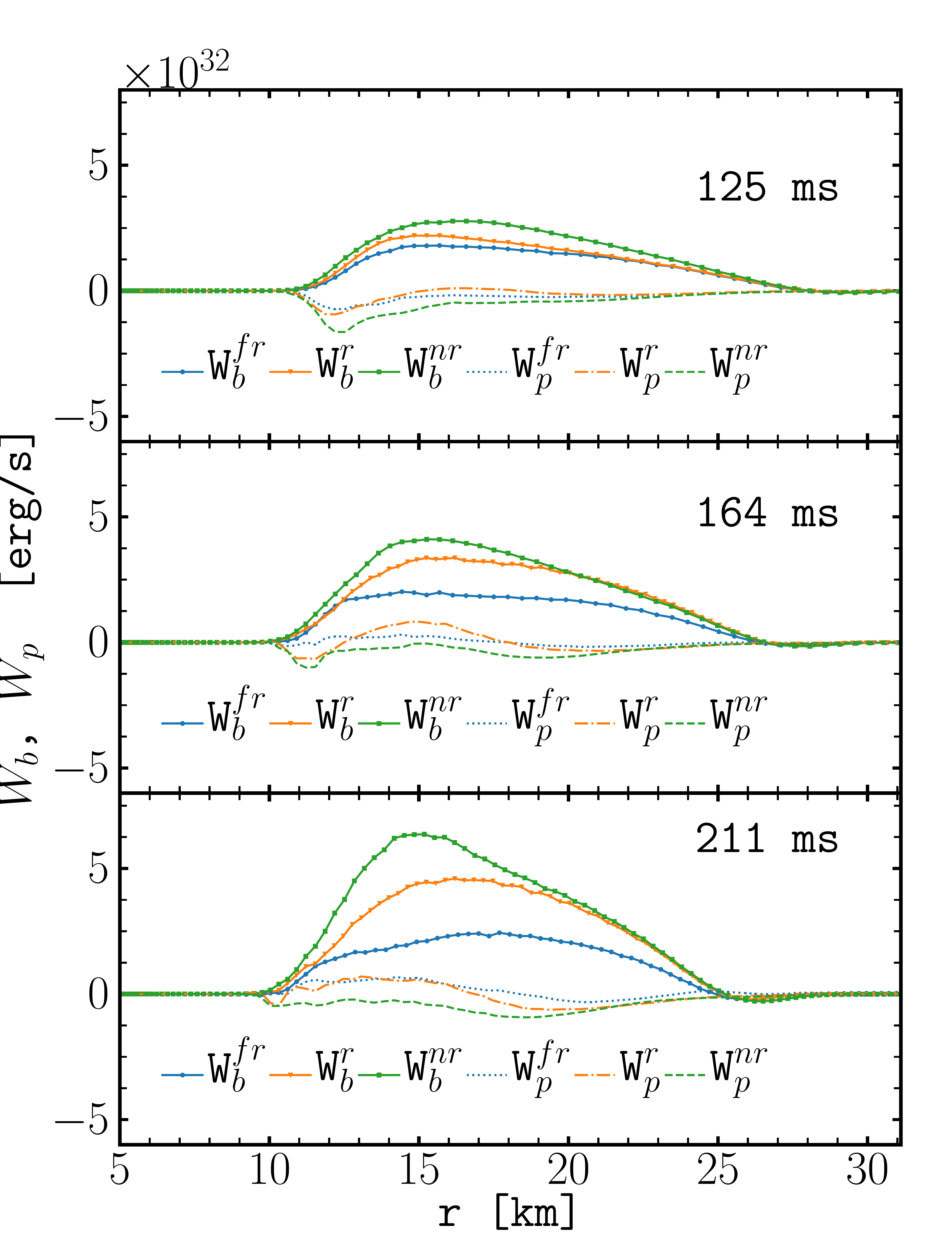}
  \caption{Angle and time averaged buoyancy work due to density
  fluctuations and work resulting from pressure fluctuations as a
  function of radius. The time averaging was done over 10 ms.  The
  times shown in the top left corner of each panel indicate the start
  of the time-window over which the averaging was performed.}
\label{figp2:fm}
\end{figure}
In \fig{figp2:fm} we show $W_b$ and $W_p$ for our three models at
three different times.  It is clear that the buoyancy work is greatest
in model m15nr and smallest in model m15fr.  The kinetic energy
injected by buoyancy forces decreases with increasing rotation rate.
Pressure fluctuations seem to be most important at the bottom of the
convective layer, where they act as a net sink of kinetic energy.

Based solely on the damping effect that rotation has on PNS
convection, we would expect an overall reduction of high-frequency GW
emission in the two rotating models. We would also expect that the
fastest rotating model m15fr emits the weakest high-frequency
signal. However, the picture is not so simple, because we also have to
consider the effects of SASI activity, whose strong influence exerted
through a coherent large-scale modulation of the accretion flow onto
the PNS effectively perturbs the PNS surface. These perturbations
reach into the interior of the PNS and also influence the convective layer
of the PNS.  

The strong downflows created by the SASI do not only excite
non-resonant g-modes in the PNS surface, but also resonant g-mode
oscillations. In model m15fr, where the spiral SASI mode dominates the
post-shock flow, we see a particularly strong low-frequency signal. This
indicates that the spiral mode can effectively perturb the PNS
surface. The same behaviour is seen in model s27 and model s20 of
\cite{andresen_17}.  The PNS surface perturbations, in turn, lead to a
stronger resonant g-mode excitation and a stronger high-frequency GW
signal.  However, because the forcing has a broad spectral
distribution, the resulting low-frequency and high-frequency GW
emission is also ``broadened'.  In model m15fr with its strong spiral
mode, GWs are emitted over a wider range of frequencies than in model
m15r.

\begin{figure}         
\centering                            
\includegraphics[width=0.4\textwidth]{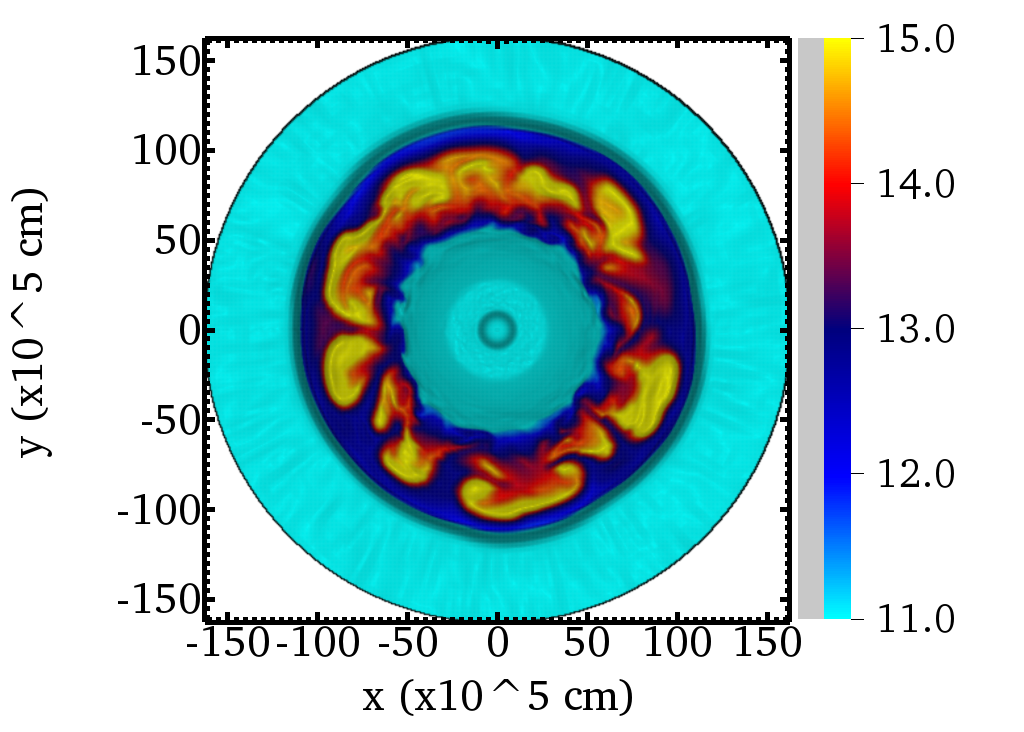} \\
\includegraphics[width=0.4\textwidth]{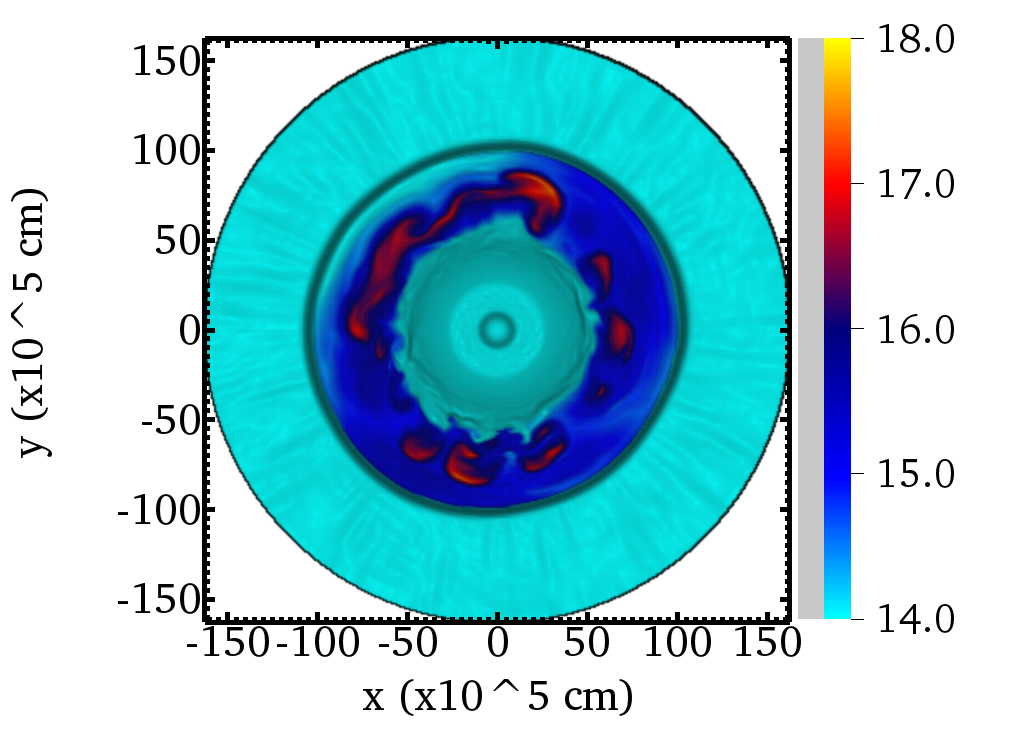} \\
\includegraphics[width=0.4\textwidth]{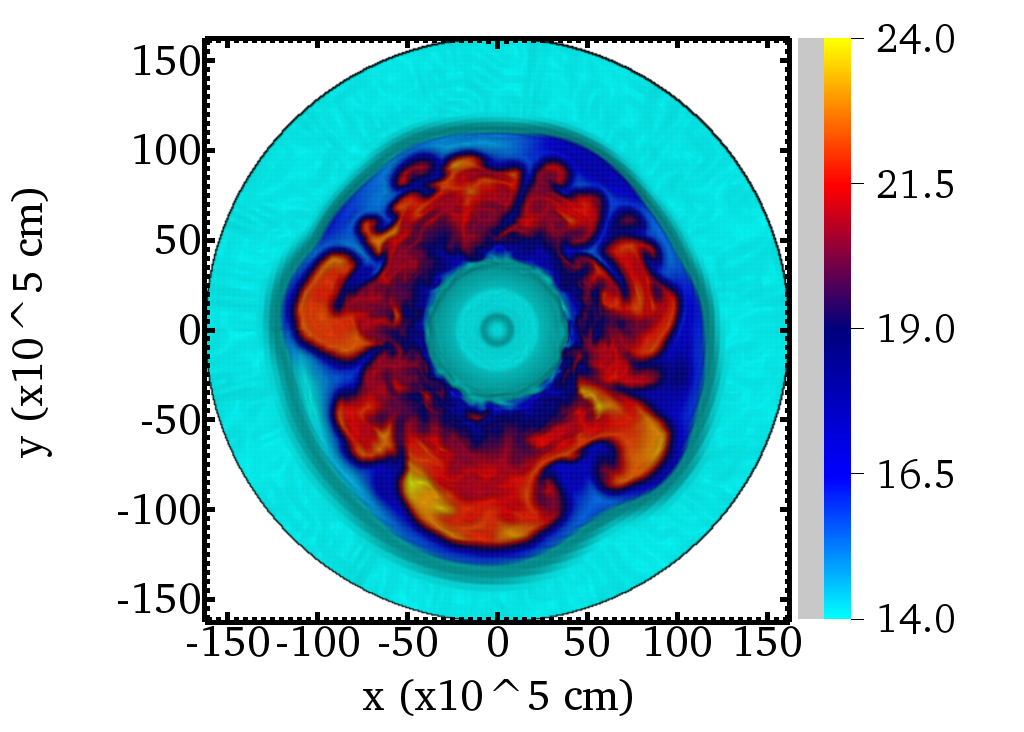}
\caption{Distribution of entropy per baryon in the equatorial plane
  for model m15r at 167 (top), 210 (middle), and 343\,ms (bottom)
  post-bounce.  The entropy is given in units of Boltzmann's constant
  $k_b$.}
\label{figp2:sto}
\end{figure}

The reduction of high-frequency GW emission generated by PNS
convection can most clearly be diagnosed in model m15r, in which the
average shock radius reaches a minimum at $\sim 200\,$ms post-bounce
after a $70\,$ms long period of recession (\fig{figp2:rsh}). The
small average shock radius favours SASI activity over neutrino-driven
convection, and we see the development of low-amplitude shock
oscillations (\fig{figp2:sasi}, middle panel). Convection, on the
other hand, is quenched.  

\fig{figp2:sto} shows the distribution of the entropy per baryon of
model m15r in the equatorial plane at three snapshots after core
bounce. In the top and bottom panels one sees the typical hot bubbles
that are characteristic of neutrino-driven convection, while in the
middle panel such bubbles are considerably suppressed (the same behaviour can be
seen in \fig{figp2:3dpics2}).  The growth of low amplitude SASI
activity and the suppression of convection in the post-shock layer
between $\sim 180$ and $\sim 250\,$ms post-bounce are reflected in
the GW signal as a weak emission at low-frequencies and a complete
absence of the high-frequency signal component.  At $\sim 250\,$ms
post-bounce the high-frequency emission sets in once more, at the same
time as the weak SASI oscillations subside.  

The SASI activity in model m15r is not strong enough to excite
resonant g-modes in the PNS, since we see high-frequency emission
during the time when there is no SASI activity, and we see no
high-frequency emission during the time when the SASI oscillations are
strongest. Moreover, the high-frequency signal vanishes when there is
very weak hot-bubble convection in the post-shock layer. It is, therefore,
clear that high-frequency emission is caused by convective plumes from
the post-shock layer impinging on the PNS surface.

The reduction of GW emitted from the PNS can also be seen in model
m15fr. After the onset of shock revival the accretion rate onto the
central object decreases and the violent downflows created by strong
SASI activity cease to exist.  At the same time the GW amplitudes
decrease strongly, indicating that excitation of surface g-modes from
above the PNS is the main source of high-frequency GW emission.  This
finding is very different from the behavior of model s20s of
\cite{andresen_17}, in which activity within the PNS increased after
the onset of shock revival and consequently led to a strong increase
of the GW signal amplitude. Thus, the nature of PNS convection,
in addition to the flow activity in the post-shock layer, is an
important factor in determining the GW signal from core-collapse
supernovae.

\begin{figure}
    \begin{center}
    \includegraphics[width=0.45\textwidth]{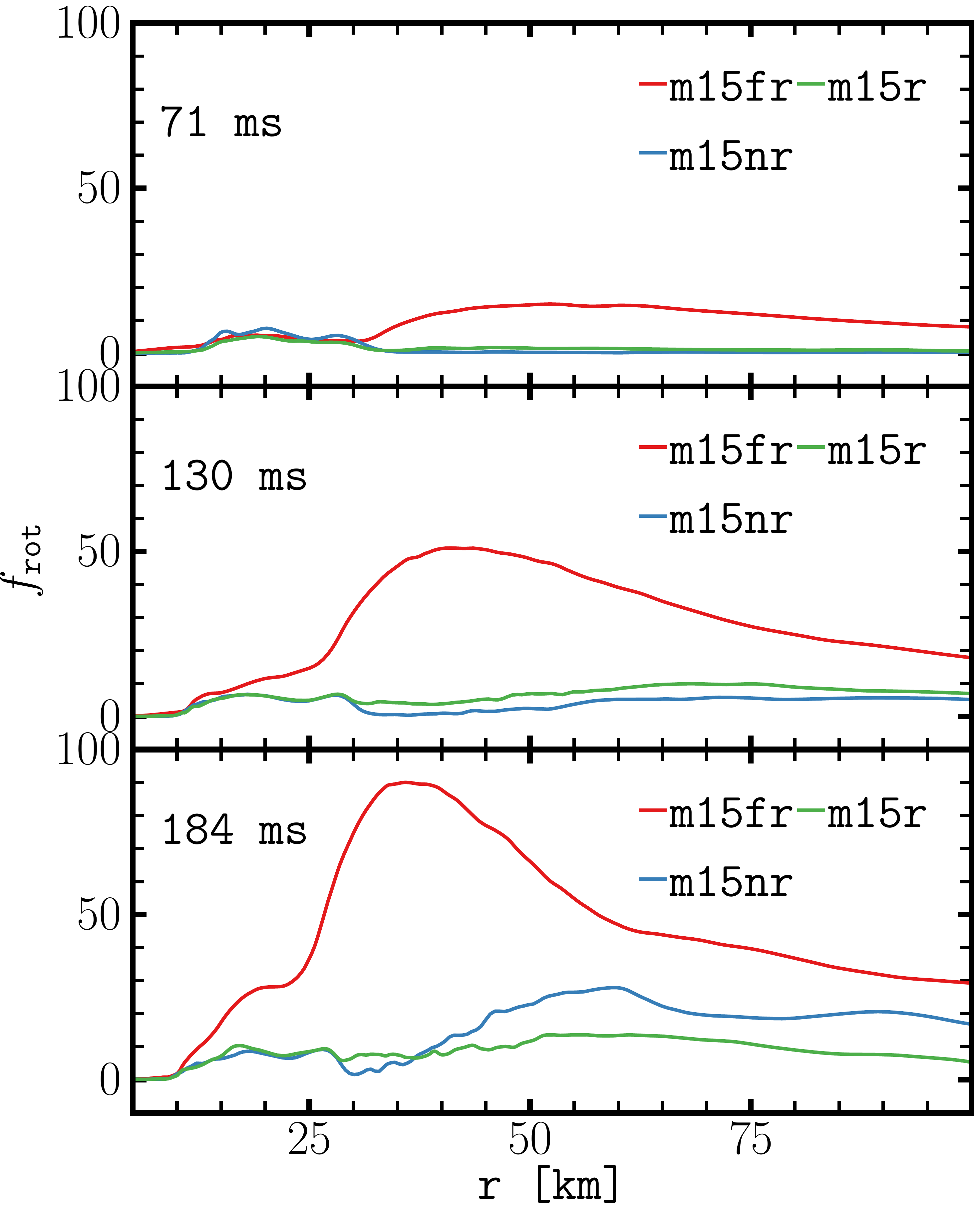}
    \caption{Rotational frequency, defined by \eq{eq:rotfreq}, 
    as a function of radius at three different times. The time is given in ms after bounce.}
    \label{fig:rotrates}
    \end{center}
\end{figure}

{The spiral SASI mode induces angular mass motions in the outer PNS, which excites low-frequency GWs.
  The initial rotation of the PNS, which is inherited from the stellar progenitor, is not sufficient
enough to have any noteworthy impact on the signals. A rotating triaxial configuration would produce GWs, but
the initial rotation of the PNS is not great enough to explain the frequency of the GWs we observe. In \fig{fig:rotrates} we
plot the angle averaged rotation frequency ($\langle f_{\mathrm{rot}} \rangle$) as a function of radius. The rotational frequency of
a fluid element around the origin is
\begin{equation} \label{eq:rot1}
  f_{\mathrm{rot}} = \frac{|\vec{r} \times \vec{v}|}{2\pi r^2}.
\end{equation}
Here $\vec{r}$ is the position vector and $\vec{v}$ the velocity vector.
In spherical coordinates the angle average of \eq{eq:rot1} is given by
\begin{equation} \label{eq:rotfreq}
\langle f_{\mathrm{rot}} \rangle = \frac{1}{4\pi}\int \frac{1}{2\pi r} \sqrt{{v_\theta}^2 + {v_\phi}^2} d\Omega,    
\end{equation}
where ${v_\theta}$, and ${v_\phi}$ are the $\theta$ and $\phi$ components of the velocity vector in spherical coordinates, respectivly.
In the first panel of \fig{fig:rotrates} we see that the rotation rate is initially too low
to explain the GW emission. Only later, when the SASI becomes active, are the outer PNS layers spun up sufficiently
to account for the low-frequency signal.
This argument is also supported by the fact that the low-frequency emission is
mainly seen during the SASI phases. The mass motions in the PNS surface, induced by the SASI, continue after the instability
subsides. The continued mass motions in the PNS surface are the source of the weak low-frequency emission seen in the spectrograms
after periods of strong SASI activity.

In the last panel of \fig{fig:rotrates} we can see that the average rotation frequency (in the post-shock layer)
is greater in model m15nr than in model m15r. The strong SASI activity that develops in model m15nr
induces angular mass motions, which at later times causes the post-shock layer fluid of model m15nr to rotate faster than
the corresponding fluid in model m15r.}

\section{Decrease in the central SASI frequency}
\label{sec:sasi2rsh}
{In the spectrogram (\fig{figp2:spec}) of model m15fr we can clearly see a decrease in
the central SASI frequency after the shock starts to expand. Normally shock expansion
leads to rapid decay of SASI activity, but in our fastest rotating model
rotation supports the SASI for a few cycles even after shock expansion has started.
This does not happen in the non-rotating and exploding model of \cite{andresen_17}.
Since the SASI frequency is connected to the average shock radius, we can actually
probe the shock expansion using the low-frequency GW emission. More importantly,
if we were to observe such an effect it would be a strong indicator that
the core of the progenitor was rotating.

The typical frequency of GWs produced by the SASI ($f^{\text{SASI}}_{\text{GW}}$) can be estimated as follows
\begin{equation} \label{eq:fsasi}
  f^{\text{SASI}}_{\text{GW}} = \frac{2}{\tau_{\text{SASI}}} \approx
  2\Bigg[ \int_{R_{\text{PNS}}}^{R_{\text{S}}}\ud r \bigg( \frac{1}{c_s} + \frac{1}{|v_r|}\bigg)  \Bigg]^{-1}.
\end{equation}
In the above expression $\tau_{\text{SASI}}$ is the typical SASI oscillation period,
$R_{\text{PNS}}$ is the radius of the PNS, $R_{\text{S}}$ represents the average shock radius,
$c_s$ is the sound speed, and $v_r$ is the typical radial accretion
velocity (the advection velocity). We refer to \cite{foglizzo_07},
\cite{scheck_08}, \cite{mueller_14}, and \cite{janka_17} for more details
about the typical SASI time-scale and remind the reader that the factor of
two comes from frequency doubling due to the quadrupole nature of GWs.

The advection velocity in the post-shock region is, to first order, a linear function of the radius and given by
the post-shock velocity $v_{\text{PS}} = -\beta^{-1} \sqrt{G M_{\text{PNS}}/R_{\text{S}}}$ ($\beta$ is the ratio of
post-shock to pre-shock density and $M_{\text{PNS}}$ is the mass of the PNS) as $v_r = v_{\text{PS}} r / R_{\text{S}}$.
Since advection the velocity is typically much smaller than the sound speed, \eq{eq:fsasi} can be approximately
be written as
\begin{equation} \label{eq:fsasis}
  f^{\text{SASI}}_{\text{GW}} \approx 2\frac{v_{\text{PS}}}{R_{\text{S}}} \bigg[ \ln \big(R_{\text{S}}/R_{\text{PNS}}\big)\bigg]^{-1}.
\end{equation}
This means that the ratio of $f^{\text{SASI}}_{\text{GW}}$ at two different times should be
roughly equal to the inverse of the ratio of the average shock radii
at the two respective times. (Actually, because of the dependence of $v_{\text{PS}}$ on $R_{\text{S}}$ and the logarithmic
factor the functional variance with $R_{\text{S}}$ is a bit steeper).
In model m15fr $f^{\text{SASI}}_{\text{GW}}$ shrinks roughly by
a factor of two, from $\sim 100\,$Hz to $\sim 40\,$Hz, during the
$100\,$ms long time window between $\sim 150\,$ and $\sim 250\,$ms after
core bounce. During the same time window the average shock radius increases
from approximately 180 km to 440 km, in other words a factor of around two
(see \fig{figp2:spec} and \fig{figp2:rsh}).
In principle this effect should be visible whenever the shock expands or retreats,
but the effect is normally not important because of the relatively small changes
in the average shock radius during the SASI dominated phase. However, since the
SASI, in model m15fr, persists for a few cycles after shock expansion sets in
we can see this effect clearly. In contrast, in the non-exploding models m15nr and m15r,
the gradually receding shock front leads to a slowly increasing frequency ($f^{\text{SASI}}_{\text{GW}}$) of the
low-frequency component of the GW emission.

To fully understand how much we can learn about rotation from observing such
an effect, it will be necessary to perform several simulations with a wider
range of rotation rates and initial conditions and to
better understand how the SASI behaves in a rotating medium.}
%%%%%%%%%%%%%%%%%%%%%%%%%%%%%%%%%%%%%%%%%%%%%%%%%%%%%%%%%%%%%%%%%%%%%%%%%%%%

%%%%%%%%%%%%%%%%%%%%%%%%%%%%%%%%%%%%%%%%%%%%%%%%%%%%%%%%%%%%%%%%%%%%%%%%%%%%
\section{The standing accretion shock instability, rotation, and
  resolution}
\label{sec:sasi}
For the models considered here, there is no clear correlation between
the development of SASI activity and the initial rotation rate of the
progenitor. Model m15r, in which the rotation rate and profile is in
accord with stellar evolution calculations, does not develop strong
SASI activity. On the other hand, both the fast rotating model m15fr
and the non-rotating model m15nr develop strong SASI activity.

How exactly rotation influences the SASI growth rate and saturation
does not seem to be a simple function of progenitor rotation rate.
Recently \cite{blondin_17} studied the effects of rotation by means of
idealised hydrodynamic simulations of a standing accretion shock (in
2D and 3D).  The results of their study are in good agreement with the
perturbative study of \cite{yamasaki_08}, who found that the linear
growth rate of non-axisymmetric SASI modes is an increasing function
of the progenitor rotation rate.  However, in the non-linear regime
\cite{kazeroni_17} do not find a monotonic connection between the
rotation rate and the saturation amplitude of the SASI. In fact, Fig.5
of \cite{kazeroni_17} indicates that SASI activity may decrease with
increasing rotation rate, at least at low to moderate rotation rates.
{It should be noted that the models of \cite{kazeroni_17} 
are idealised simulations which do not include the same physics  
as the models our work is based on. This leads to different conditions in the post-shock volume and we
must, therefore, be careful when extrapolating results from such studies to the models discussed in our work.}

An additional complication comes from the fact that model m15nr was
simulated with half the angular resolution of the other two models.
Lower angular resolution has been found to favour the growth SASI activity,
because energy accumulates at larger scales and parasitic instabilities
(Rayleigh-Taylor, Kelvin-Helmholtz), which tap energy from SASI motions,
are suppressed  \citep{hanke_12}.
\cite{abdikamalov_15} found the opposite, they concluded
that decreasing the spatial resolution damps SASI oscillations.
However, their simulations were performed with a Cartesian grid and their
results could therefore be a consequence of changing the radial resolution in
addition to the angular resolution (see \cite{sato_09}). 
Independent of whether rotation suppresses the growth of SASI activity
in the slowly rotating model m15r or whether SASI activity is enhanced
in the non-rotating model m15nr because of insufficient
angular resolution, the absence of strong SASI activity in model m15r
reduces the overall strength of the GW signal.

%%%%%%%%%%%%%%%%%%%%%%%%%%%%%%%%%%%%%%%%%%%%%%%%%%%%%%%%%%%%%%%%%%%%%%%%%%%%
\section{Core bounce signal} \label{sec:cb}
\begin{figure}         
\centering                            
\includegraphics[width=0.45\textwidth]{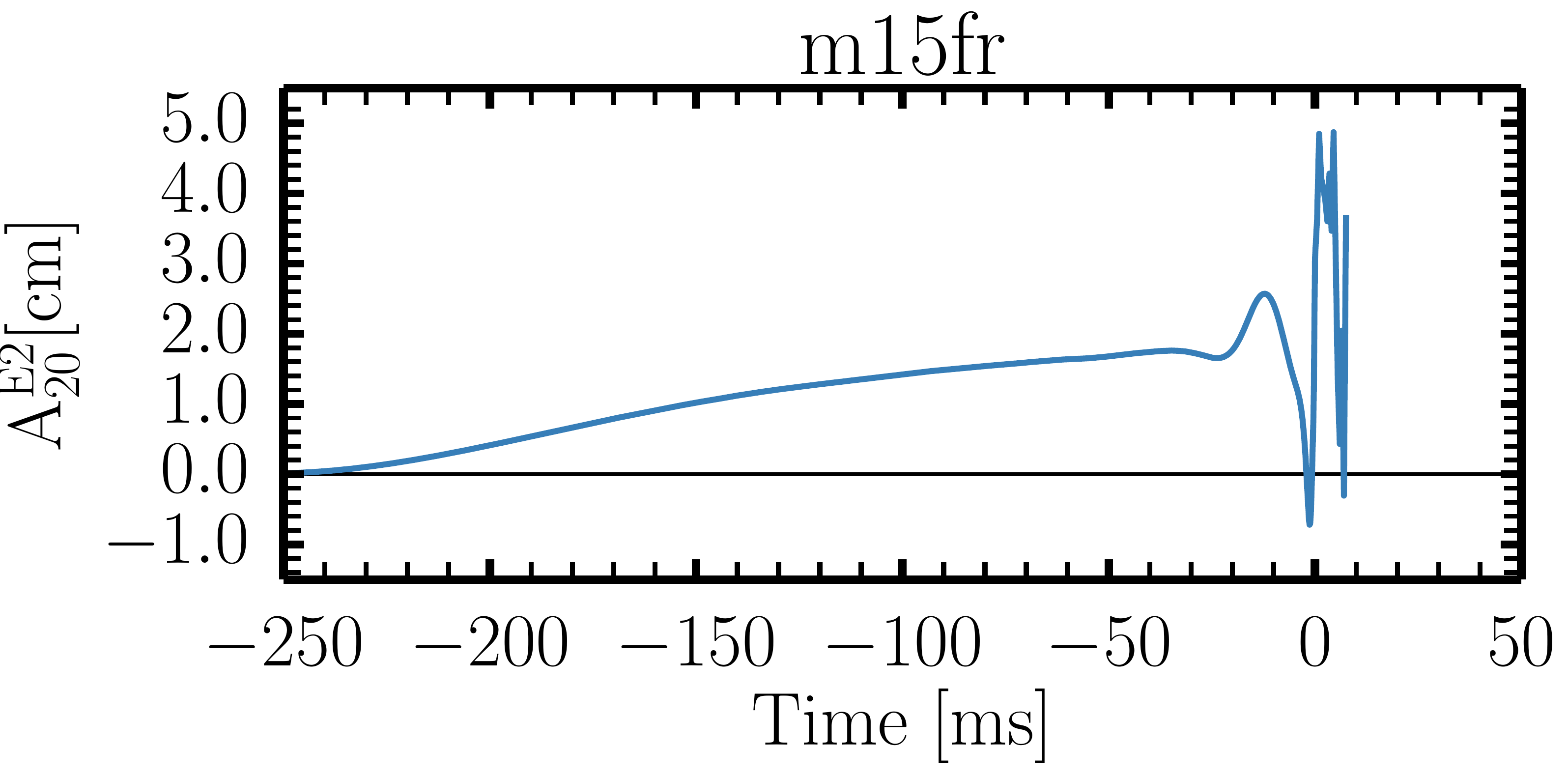}\\
\includegraphics[width=0.45\textwidth]{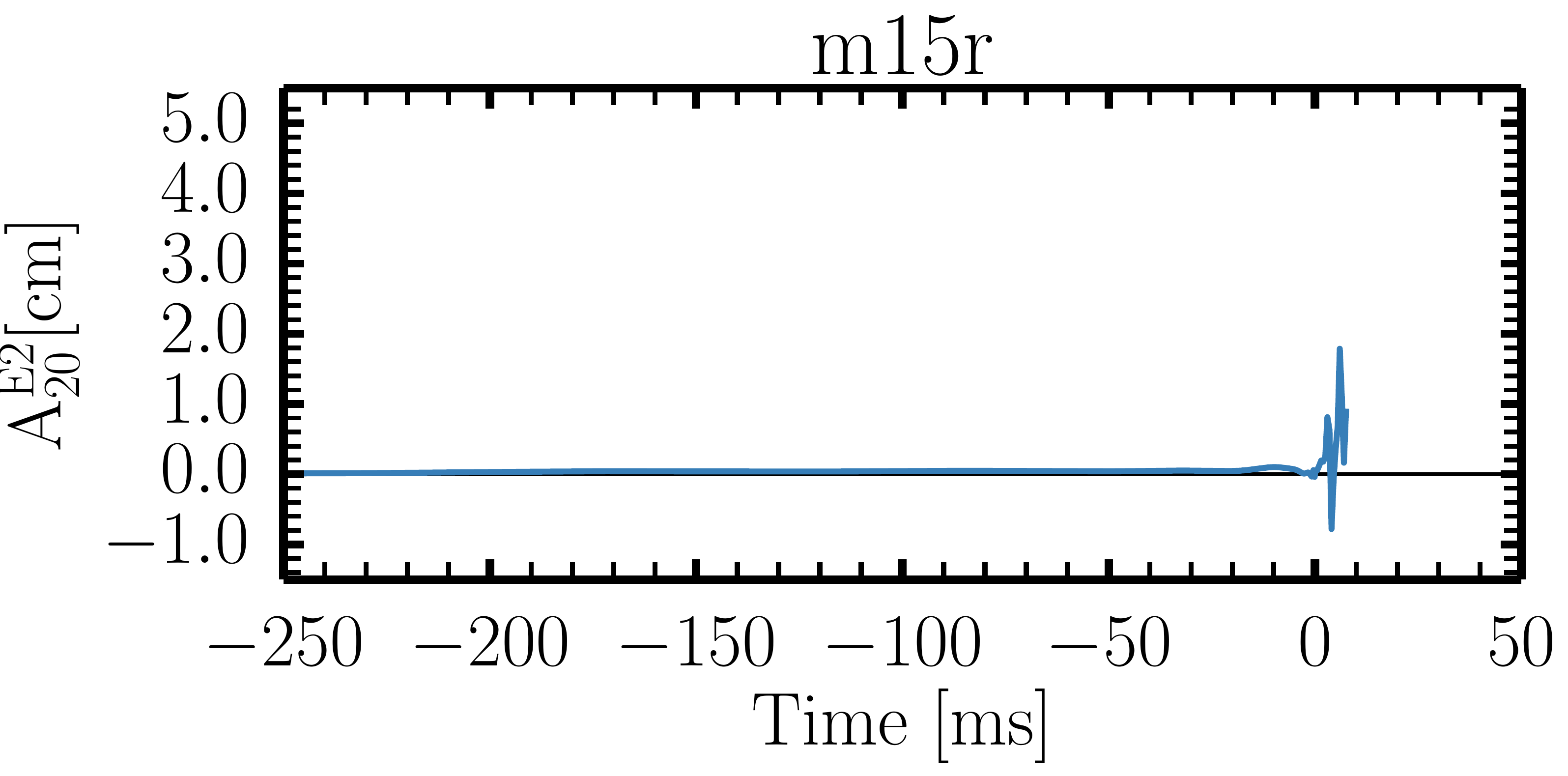}
\caption{GW amplitudes from 2D simulations of collapse and bounce of
  models m15fr (top), and m15r (bottom). The amplitudes are calculated
  according to \eq{eq:2dquad}.}
\label{figp2:bounce}
\end{figure}

Because the 3D simulations of \citet{Summa_18} only start
approximately 10\,ms after core bounce, the expected GW signal
associated with core bounce is not present in our waveforms in
\fig{figp2:amps}.  The evolution of the models from the onset of
core-collapse until the start of the 3D simulations was performed by
\citet{Summa_18} in 2D, and the resulting GW amplitudes are shown in
\fig{figp2:bounce}.  The amplitudes (of the two rotating models) are
calculated according to \eq{eq:2dquad}.

The flattening of the core that occurs during collapse causes a
positive and steadily increasing GW amplitude, in particular in model m15fr.
The abrupt halt of the collapse, followed by the expansion of the core leads to a sudden
increase in the GW amplitude and a subsequent sharp drop towards
negative values. After bounce the waveforms show the typical large
oscillations associated with the ring-down of the core, see \eg
\cite{mueller_82, finn_90, moenchmeyer_91, yamada_95, zwerger_97}.
When considering the factor ${1}/{8}\sqrt{15/\pi} \approx 0.27$ in
\eq{eqT:htht}, the bounce signal of the two models is slightly (model
m15fr) and considerably (model m15r) weaker than the emission from the
signal of the post-bounce phase.

%%%%%%%%%%%%%%%%%%%%%%%%%%%%%%%%%%%%%%%%%%%%%%%%%%%%%%%%%%%%%%%%%%%%%%%%%%%%
\section{Detection prospects} \label{sec:det}
\begin{table*}
  \caption{Signal-to-noise ratios for models m15fr, m15r, and 
    m15nr for three different frequency domains: $20\ldots 250\, \mathrm{Hz}$ (low), 
    $250\ldots 1200\, \mathrm{Hz}$ (high), and $20 \ldots 1200\, \mathrm{Hz}$ (total). 
    The table gives values for AdvLIGO and the Einstein Telescope. For
    the latter, we calculate the SNR for two different modes of
    operation (ET-B and ET-C) assuming a source at a distance of $10\,\mathrm{kpc}$.   \label{tablep2:SNR}} 
  \begin{tabular}{c|c|c|c|c|c|c|c|c|c|c|c|c}
    \toprule
    \multicolumn{1}{l|}{} & \multicolumn{4}{c}{m15fr} & \multicolumn{4}{c}{m15r} & \multicolumn{4}{c}{m15nr} \\ 
    \multicolumn{1}{l|}{}   & Low   & High & Total & Low/High & Low  & High & Total & Low/High & Low  & High & Total & Low/High \\ \toprule 
    \multicolumn{1}{l|}{AdvLIGO} & 10.8  & 4.9  & 11.9  & 2.20     & 2.6  & 2.4  & 3.5   & 1.08     & 3.5  & 4.3  & 5.5   & 0.81        \\ 
    \multicolumn{1}{l|}{ET-C}    & 133.2 & 67.6 & 149.4 & 1.97     & 32.2 & 34.4 & 47.1  & 0.93     & 46.5 & 59.3 & 75.2  & 0.78        \\
    \multicolumn{1}{l|}{ET-B}    & 224.6 & 83.5 & 239.6 & 2.69     & 53.7 & 41.1 & 67.7  & 1.30     & 74.0 & 72.0 & 103.2 & 1.02        \\ \toprule 
  \end{tabular}
\end{table*}

Following the procedure laid out in \cite{andresen_17}, we use the
signal-to-noise ratio (SNR) (see for example \cite{flanagan_98a}) for
a matched filtered signal to provide a rough assessment of the
detectability of the GW signals described in this work.  Under the
asumptions of an optimally orientated detector and that the emission
is more or less isotropic the SNR for a matched filter signal is given
by
\begin{equation}
\label{eq:snr}
(\mathrm{SNR})^2
=
4 \int_0^\infty df \frac{|\tilde{h}(f)|^2}{S(f)}
=
\int_0^{\infty} df \frac{h_c ^2}{f^2 S(f)},
\end{equation}
where
\begin{equation}
\label{eq:h_c}
h_c =\sqrt{ \frac{2 G}{\pi^2 c^3 D^2} \frac{d E}{df}}
\end{equation}
is the characteristic strain and $S(f)$ is the power-spectral density
of the detector noise as a function of frequency $f$. The assumption
of isotropic emission alows us to replace $\tilde{h}(f)$ in
\eq{eq:snr} with $h_c$.  In reality $\tilde{h}(f)$ will depend on
the observer position, but for the purpose of rough estimates these
variations are small enough to justify the assumption of isotropic
emission. We can, therefore, express the SNR in terms of the GW energy
spectrum, which is calculated according to \eq{eq:enspc}

We will calculate the SNR in three frequency bands, these bands are
the same that were used by \cite{andresen_17}. This allows us to
easily compare the results and gauge the effect of rotation. It is
particularly interesting to determine whether or not excess of power
at low frequencies remains a fingerprint of strong SASI activity in the
models presented here, as has been found to be the case for
non-rotating 3D models \citep{kuroda_16,andresen_17}.  The three bands
are defined by the following frequency ranges:
$20\ldots 250 \, \mathrm{Hz}$, $250\ldots 1200 \, \mathrm{Hz}$, and
$20 \ldots 1200 \, \mathrm{Hz}$. These three bands together with \eq{eq:snr} define
\begin{align}
(\mathrm{SNR}_\mathrm{low})^2 &= \int_{20}^{250} df \frac{h_c ^2}{f^2 S(f)}, \nonumber \\
(\mathrm{SNR}_\mathrm{high})^2 &= \int_{250}^{1200} df \frac{h_c ^2}{f^2 S(f)}, \nonumber \\
 \text{and} \ (\mathrm{SNR}_\mathrm{total})^2 &= \int_{20}^{1200} df \frac{h_c ^2}{f^2 S(f)}.
\end{align}

We calculate the SNR for the zero-detuning-high power configuration of
Advanced LIGO \citep{adv_sens}, and the B \citep{et_b} and C
\citep{et_c} configuration for the Einstein telescope. These
configurations are referred to as as AdvLIGO, ET-B and ET-C.

From table~\ref{tablep2:SNR}, it is clear that the SNR does not increase
with the initial rotation rate of the progenitor. The trend is rather
that the SNR correlates with SASI activity. Model m15fr that develops the
strongest SASI activity also has the highest SNR, in all bands and for
all detectors. Our non-rotating model (m15nr) also develops strong
SASI activity and shows higher SNR than model m15r. It should be
pointed out that model m15nr, despite its lower angular resolution,
is not largely different from the other non-rotating models investigated by
\cite{andresen_17}.
In model m15r we see the lowest SNR values and only weak SASI activity. However, the ratio of
$\mathrm{SNR}_\mathrm{low}/\mathrm{SNR}_\mathrm{high}$ is greater for
model m15r than for model m15nr. This happens because of the reduced
high-frequency emission in model m15r due to dampening of
PNS convection by rotation. It is not strong low-frequency emission
in model m15r that gives rise to the high ratio of
$\mathrm{SNR}_\mathrm{low}/\mathrm{SNR}_\mathrm{high}$, high compared
to model m15nr, but rather the weak high-frequency emission.  In model
m15fr the effect is opposite, strong low-frequency emission paired
with reduced high-frequency emission gives high ratios of
$\mathrm{SNR}_\mathrm{low}/\mathrm{SNR}_\mathrm{high}$ in all the
three detectors. This means that excess energy in the low-frequency
band is not such a clear indication of strong SASI activity as
initial 3D models suggested.

Another unfortunate consequence of rotation, unless rapid, is the possibility of greatly
reduced signal strength. PNS convection and SASI activity
are the main drivers of GW emission and rotation can quench both of
these processes. As opposed to rapid rotation, moderate rotation can
make the signal harder to detect. On the other hand, model m15fr
develops a very strong spiral SASI mode and this enhances the
detectability of this model. The exact dependence of the growth of
SASI activity and initial progenitor rotation will be crucial for
predicting the strength of the GW signal.

\subsection{Future improvements}
{The back of the envelope estimates presented in our
  work are based on detecting an excess of power in given frequency bands.
  Our results are in good agreement with standard search methods
  \citep{gossan_16,ligo_sn_search,powell_17} that rely on excess power in
  the detectors during an astrophysically motivated time period.
  Currently, it seems that the detection of GWs
  from core-collapse supernovae should not be expected beyond a few kpc.
  However, signal recovery methods can be improved by incorporating known
  information about the expected signal into the analysis. While the time
  domain GW signal is stochastic there are clear patterns in the spectrograms (\fig{figp2:spec}).
  The time-frequency evolution of the high-frequency emission is one example of
  a distinct and robust signal feature. By preferentially searching for
  signal contributions characterized by phenomenological time-frequency tracks
  motivated by theoretical predictions, the detection prospects may be improved.
  Third-generation interferometers will possess noise floors with about one order of
  magnitude improvement in the amplitude of the design Advanced-LIGO noise floor,
  which is already a factor of two better than the second LIGO-Virgo
  \citep{ligo_noise,ligo_3gnoise} observing run.
  The improved sensitivity of the next generation of detectors and advancements in data analysis
  should extend the detection horizon of GWs from core-collapse supernovae well
  beyond the estimates presented in this work.}

%%%%%%%%%%%%%%%%%%%%%%%%%%%%%%%%%%%%%%%%%%%%%%%%%%%%%%%%%%%%%%%%%%%%%%%%%%%%
\section{Conclusion and discussion} \label{sec:con}
In this work, we studied how progenitor rotation in 3D models affects the GW signal
of core collapse supernovae.  We have not studied very rapidly rotating
models, but rather focused on the regime of more moderate progenitor
rotation.  Our main findings are:

\begin{enumerate}
\item Moderate rotation does not change the frequency structure of the
  GW signal, compared to the signal from non-rotating models.  We see
  the familiar two-component structure with a high-frequency and a
  low-frequency signal component \citep{andresen_17}.
\item We find that the high-frequency emission instigated by PNS
  convection is weaker in the rotating models because rotation has a
  stabilising effect on PNS convection and decreases the amount of
  energy dissipated in the overshooting region of the PNS.  This
  becomes particularly apparent in model m15r, where we find a strong
  reduction of the GW amplitudes during a period of time when
  post-shock convection is weak. The generally weak amplitudes in this
  model, which has no strong SASI activity, reaffirm the fact that
  post-shock convection is a weak source
  of GW excitation, compared to PNS convection and SASI activity.
\item Of the three models presented in this chapter the fastest
  rotating model emits the strongest GW signal because it develops the
  strongest spiral SASI mode. Based on the models presented here one
  should not conclude that the strength of the signal will increase
  with increasing progenitor rotation rate. The conclusion should
  instead be that the stronger the spiral mode of the SASI is the
  larger are the amplitudes of the GW signal. However, it remains
  unclear whether faster rotation in a monotonic dependence leads
  to stronger SASI activity.
\item It should be emphasised that model m15r, where the rotation rate
  is exactly in accordance with stellar evolution calculations, shows
  the weakest GW signal.
\item Unlike model s20s of \cite{andresen_17}, the GW signal of model
  m15fr decreases after the onset of shock revival. This reduction is
  due to the small contribution to the total signal from mass motions
  instigated by PNS convection.
\end{enumerate}

Prior to this work, the GW signals of rapidly rotating models have
been studied by several authors
\citep{mueller_82,rampp_98,shibata_05,ott_05,scheidegger_10,kuroda_14,takiwaki_16}.
A common feature of these studies is that they tend to predict rather
strong emission of gravitational radiation.  During the post-bounce
phase, rapid rotation can lead to the development of novel flow
patterns that are not observed in slowly/non-rotating models.
\cite{rampp_98} and \cite{shibata_05} found that very rapid rotation
can lead to a bar-like deformation of the central core. In the
somewhat slower rotating models of \cite{ott_05}, \cite{kuroda_14},
and \cite{takiwaki_16} the development of a low-mode spiral
instability (low-T/W) was found. 
{While the rotation rate of model m15fr
exceeds the threshold found in the simplified models of
\cite{kazeroni_17}, where a co-rotation instability can develop,
we found no signs of the low-T/W instability in our models \citep{Summa_18}. It is not clear
that exceeding this threshold guarantees the development of the spiral mode \citep{foglizzo_17_CNUT}.
The results from idealised studies should be applied to
our models only with caution.}

These asymmetric and rapidly rotating
structures lead to strong GW emission.  It is, however, not likely
that a large fraction of the core collapse supernova progenitors have
rapidly rotating, or even moderately rotating, cores. Observations of
pulsars put strong constraints on the rotation rate of core-collapse
progenitors. It has been estimated that most pulsars are formed with
rotation periods of a few tens to hundreds of milliseconds
\citep{vranesevic_04,popov_12,noutsos_13}.  The recent study by
\cite{kazeroni_17} concludes that the one armed-spiral instability
\citep{ott_05,kuroda_14,takiwaki_16} is not able to spin down the PNS
enough to make rapidly rotating progenitors compatible with the spin
rate of young pulsars. Additionally, stellar evolution models which
include the effects of magnetic fields predict slowly rotating stellar
cores \citep{heger_05}. Results from asteroseismology
\citep{beck_12,mosser_12} indicate that the cores of low-mass red
giants rotate slower than what is expected from stellar evolution
calculations \citep{cantiello_14,deheuvels_14}.  According to results
from asteroseismology, angular momentum loss due to stellar winds
seems to play a bigger role than currently predicted by stellar
evolution calculations \citep{cantiello_14}.

The rotation rates of the two rotating models studied here are more
along the lines of what is expected from state of the art stellar
evolution calculations \citep{heger_05} and observations
\citep{beck_12,mosser_12,popov_12,noutsos_13,cantiello_14,deheuvels_14}.
We find no qualitative difference in the GW signals from the rotating
models, compared to those from the non-rotating models in \cite{andresen_17}.
We must, therefore, conclude that a stochastic signal with amplitudes of a few
centimetres seems to be the generic core-collapse GW signal. Hence,
moderate to low progenitor rotation will not significantly increase
the detectability of GWs from core-collapse supernovae. Rotation can
actually make it more difficult to detect the GW signal, because we find that
rotation decreases the signal emitted due to PNS convection in model
m15r and consequently makes the model harder to detect.

The fact that model m15nr, the non-rotating one, has an angular
resolution that is two times lower (four versus two degrees) than that
of the two other models makes it difficult to draw strong conclusions
about how the low-frequency signal changes with increasing
rotation. It is not clear why model m15nr develops strong SASI
activity and model m15r does not. Previously, lower angular resolution
in polar coordinates has been found to favour \citep{hanke_12} and lower
Cartesian resolution to suppress
\citep{abdikamalov_15} the development of strong SASI activity. {At the
same time, it is also possible that rotation quenches the growth of
SASI activity in model m15r.} This is an issue
that will have to be resolved by a more systematic study, where the
rotation rate and grid resolution are varied independently to discriminate the
impact of the individual effects. Since model m15nr, despite its lower resolution,
does not behave dissimilar from the non-rotating models described in \cite{andresen_17},
it is possible that a considerable stochastic element plays a role.
Another weakness of our study is
that the three models all start from a spherically symmetric
progenitor model.  However, it is not realistic to expect the
progenitor stars to be perfectly spherically symmetric objects.  In
fact, it has been found that asymmetries in the convective burning shells of the
progenitor can influence the shock dynamics and even help to ensure a
successful explosion
\citep{burrows_96,fryer_04,arnett_11,couch_13,mueller_15a,mueller_17}. Inhomogeneities
in the stellar core could lead to a sizable emission of GWs during the
collapse and right after core bounce. Thus, the long period of
quiescence after bounce could be an artifact of the usage of spherically
symmetric progenitors.

%%%%%%%%%%%%%%%%%%%%%%%%%%%%%%%%%%%%%%%%%%%%%%%%%%%%%%%%%%%%%%%%%%%%%%%%%%%%
\section{Acknowledgements}
We thank A.\ D\"oring for the visualizations of
Figures 2 and 3. We also thank T.\ Melson for helpful discussions about the
underlying hydro-data.
This work was supported by the Deutsche 
Forschungsgemeinschaft through SFB~1258 ``Neutrinos and
Dark Matter in Astro- and Particle Physics (NDM)'' and the
Excellence Cluster Universe EXC~153, and by the European
Research Council through ERC-AdG No.\ 341157-COCO2CASA.
The supernova simulations were performed using high-performance
computing resources (Tier-0) on SuperMUC at the 
Leibniz Supercomputing Centre (LRZ) provided by the Gauss Centre 
for Supercomputing (GCS@LRZ; LRZ project ID: pr74de).
\bibliography{p2ewm}

\end{document}